\RequirePackage{fix-cm}
\documentclass[smallcondensed]{article} %
\usepackage{amsmath}
\usepackage{amssymb}
\usepackage{amsfonts}
\usepackage[T1]{fontenc}
\usepackage{graphicx}%

\newcommand{\pa}[1]{\left( #1 \right)}
\newcommand{\br}[1]{\left[ #1 \right]}
\newcommand{\ac}[1]{\left\{ #1 \right\}}

\newcommand{\la}{\lambda}

\newcommand{\p}{\rho}
\newcommand{\s}{\sigma}

\newcommand{\q}{\theta}

\newcommand{\RR}{\mathbb{R}}
\newcommand{\CC}{\mathbb{C}}
\newcommand{\Ss}{\mathbb{S}}
\newcommand{\rarrow}{\rightarrow}

\newcommand{\rank}{\operatorname{rank}}

\newcommand{\tr}{\operatorname{tr}}

\newcommand{\pr}{\operatorname{pr}}

\newcommand{\ie}{\textit{i.e.}\ }

\newcommand{\del}{\partial}

\newtheorem{prop}{Proposition}

\newenvironment{aleq}{\begin{equation}\begin{aligned}}{\end{aligned}\end{equation}}
\newenvironment{aleq*}{\begin{equation*}\begin{aligned}}{\end{aligned}\end{equation*}}
\newenvironment{gaeq}{\begin{equation}\begin{gathered}}{\end{gathered}\end{equation}}
\newenvironment{gaeq*}{\begin{equation*}\begin{gathered}}{\end{gathered}\end{equation*}}
\newenvironment{eqe}{\begin{equation}}{\end{equation}}

\begin{document}

\title{Multimode solutions of first-order elliptic quasilinear systems obtained from Riemann invariants
}

\date{}
\maketitle
\ \vspace{-1in}\\
\begin{center}{Alfred Michel Grundland\\
              Centre de Recherche Math\'ematiques, Universit\'e de Montr\'eal,\\
                C.P. 6128, Succc. Centre-ville, Montr\'eal, (QC) H3C 3J7, Canada\\
and D\'epartement de Math\'ematiques et Informatiques,\\
 Universit\'e
du Qu\'ebec,\\
 Trois-Rivi\`eres (QC) G9A 5H7, Canada,\\
              grundlan@crm.umontreal.ca\\
and\\
Vincent Lamothe \\
              D\'epartement de Math\'ematiques et Statistique, Universit\'e de Montr\'eal,\\
                C.P. 6128, Succc. Centre-ville, Montr\'eal, (QC) H3C 3J7, Canada,\\
            lamothe@crm.umontreal.ca\\
} \end{center}

\begin{abstract}
Two new approaches to solving first-order quasilinear elliptic systems of PDEs in many dimensions are proposed. The first method is based on an analysis of multimode solutions expressible in terms of Riemann invariants, based on links between two techniques, that of the symmetry reduction method and of the generalized method of characteristics. A variant of the conditional symmetry method for constructing this type of solution is proposed. A specific feature of that approach is an algebraic-geometric point of view, which allows the introduction of specific first-order side conditions consistent with the original system of PDEs, leading to a generalization of the Riemann invariant method for solving elliptic homogeneous systems of PDEs. A further generalization of the Riemann invariants method to the case of inhomogeneous systems, based on the introduction of specific rotation matrices, enables us to weaken the integrability condition. It allows us to establish a connection between the structure of the set of integral elements and the possibility of constructing specific classes of simple mode solutions. These theoretical considerations are illustrated by the examples of an ideal plastic flow in its elliptic region and a system describing a nonlinear interaction of waves and particles. Several new classes of solutions are obtained in explicit form, including the general integral for the latter system of equations.
%\subclass{35B06 \and 35F50 \and 35F20}
\end{abstract}

Keywords:symmetry reduction method, generalized method of characteristics, Riemann invariants, multimode solutions
Mathematics Subject Classification (2000): {35B06; 35F50; 35F20}
\section{Introduction}
\label{intro} Riemann waves represent a very important class of
solutions of nonlinear first-order systems of partial differential equations (PDEs). They are
ubiquitous in the differential equations of mathematical physics,
since they appear in all multidimensional hyperbolic systems and
constitute their elementary solutions. Their characteristic feature
is that, in most cases, they are expressible only in implicit form.
For a homogeneous hyperbolic quasilinear system of first-order PDEs,
\begin{eqe}\label{eq:1}
\mathcal{A}_\alpha^{i\mu}(u)\frac{\del u^\alpha}{\del x^i}=0,\qquad i=1,\ldots,p,\quad \alpha=1,\ldots,
q,\quad \mu=1,\ldots, m
\end{eqe}%
(where $\mathcal{A}^1,\ldots,\mathcal{A}^p$ are $q\times m$ matrix functions of an unknown function $u$ and we adopt
the convention that repeated indices are summed), a Riemann wave solution is defined by the equation $u=f(r(x,u))$, where $f\colon \RR\rarrow \RR^q$, and
the function $r(x,u)=\la_i(u)x^i$ is called the Riemann invariant associated with
the vector $\lambda$. This vector $\lambda$ satisfies the equation
$\ker\pa{\la_i\mathcal{A}^i(u)}\neq 0$. Such Riemann wave solutions have rank at most
equal to one. They are building blocks for the construction of more general
types of solutions describing nonlinear superpositions of many waves
($k$-waves), which are very interesting from the physical point of
view.
\par Until now, the only way to approach this task was through the
generalized method of characteristics (GMC)
(see \textit{e.g.} \cite{Burnat:1972,CourantHilbert:1962,Jeffrey:1976,JohnKlainerman:1984,Perad:1985,Rozdestvenski:1983})
and more recently through the conditional symmetry method (CSM)
\cite{AblowitzClarkson,ConteGrundHuard:2009,Fushchych:1991,GrundHuard:2006,GrundHuard:2007,OlverVorobev1995}.
The GMC relies on treating Riemann invariants as new dependent
variables (which remain constant along the appropriate
characteristic curves of the initial system (\ref{eq:1}) and
constitute a set of invariants of the Abelian algebra of some vector
field $X_a=\xi_a^i(u)\del_x^i$ with $\la_i^a\xi_a^i=0$ for $1\leq
a\leq k<p$. This leads to a reduction of the dimension of the
problem. The most important theoretical results obtained with the
use of the GMC or CSM \cite{GrundHuard:2007} include the necessary and sufficient conditions for the existence of Riemann $k$-waves in multidimensional systems. It was shown
\cite{Perad:1985} that these solutions depend on $k$ arbitrary
functions of one variable. Some criteria were also found
\cite{Perad:1985} for determining the elastic or nonelastic
character of the superposition of Riemann waves described by
hyperbolic systems, which is particularly useful in physical
applications. In applications to fluid dynamics and nonlinear field
theory, many new and interesting results were obtained
\cite{Boillat:1965,Burnat:1972,DubrovinNovikov:1983,FerapontovKhus:2004:1,FerapontovKhus:2004:2,JohnKlainerman:1984,Mises:1958,Rozdestvenski:1983,Whitham:1974,Zakharov:1998}.
\par Both the GMC and CSM methods, like all other techniques for solving
PDEs, have their limitations. This fact has motivated the authors to
search for the means of constructing larger classes of multiple wave
solutions expressible in terms of Riemann invariants by allowing the
introduction of complex integral elements in the place of real
simple integral elements (with which the solutions of hyperbolic
systems are built \cite{Perad:1985}). This idea originated from the
work of S. Sobolev \cite{Sobolev:1934} in which he solves the wave
equation by using the associated complex wave vectors. We are
particularly interested in the construction of nonlinear
superpositions of elementary simple mode solutions, and the
proposed analysis indicates that the language of conditional
symmetries is an effective tool for this purpose. This approach is
applied to the nonstationary irrotational flow of an ideal plastic
material in its elliptic region. A further extension of the proposed method to the case of inhomogeneous systems is proposed in order to be applicable either in the elliptic or hyperbolic regions. This allows for a wider range of physical applications. The approach is based on the use of rotation matrices which obey certain algebraic conditions and allow us to write the reduced system in terms of Riemann invariants in the sense that each derivative of a dependent variable is equal to an algebraic expression (see equation (\ref{eq:ms:12})). We discuss in detail the sufficient conditions for the existence of multimode solutions. This approach is applied to a system describing the propagation of shock waves in the nonlinear interaction of waves and particles. The general integral of this system has been constructed in explicit form depending on two arbitrary functions of one variable. Through this paper, all derivatives, transformations, surfaces and solutions are assumed to be local.
\paragraph{}The organization of this paper is as follows. Section \ref{sec:2}
contains a detailed account of the generalized method of
characteristics for first-order quasilinear systems of PDEs in many
dimensions based on complex characteristic elements. In Section \ref{sec:3} we
formulate the problem of multimode solutions expressible in terms of
Riemann invariants by means of a group theoretical approach. This
allows us to formulate the necessary and sufficient conditions for
constructing these types of solutions. In Section \ref{sec:4}, the usefulness of the method developed in Section \ref{sec:3} is illustrated by the example of ideal plasticity in $(2+1)$ dimensions, in which we find several bounded solutions. Moreover, we have drawn extrusion dies and the flow inside them (limiting ourselves to the region where the gradient catastrophe does not occur).  Sections \ref{sec:5} and \ref{sec:6} comprise a new approach to solving inhomogeneous elliptic systems in order to obtain simple wave and simple mode solutions. In Section \ref{sec:7}, we show an example of a simple mode solution, obtained through the method presented in Section \ref{sec:6}. The technique is applied to a system describing the propagation of shock waves and the nonlinear interaction of waves and particles. We obtain its general solution. Section \ref{sec:8} summarizes the results obtained and contains some suggestions regarding further
developments.
%\setcounter{equation}{0}%
%################################ section 2 #####################
\section{The method of characteristics for complex integral elements}\label{sec:2}The methodological
approach assumed in this section is based on a generalized method of characteristics which has
been extensively developed (see \textit{e.g.}
\cite{Burnat:1972,DoyleGrundland:1996,GrundlandTafel:1996,Grundland:1974,Perad:1985} and references
therein) for multidimensional homogeneous and inhomogeneous systems of first-order
PDEs. A specific feature of that approach is that it has both an algebraic and geometric point of view. An
algebraization of systems of PDEs is made possible by representing the general integral elements
as linear combinations of some special elements associated with those vector fields which generate
characteristic curves in the spaces of independent variables $X$ and dependent variables $U$,
respectively (see \text{e.g.} \cite{GrundlandZelazny:1983,Perad:1985}). The introduction of these elements (called simple integral elements) turns out to be
very useful for the construction of certain classes of rank-$k$ solutions in closed form. These integral
elements correspond to Riemann wave solutions in the case of nonelliptic systems and
serve to construct multiple waves ($k$-waves) as superpositions of several single Riemann waves.
\paragraph{}The generalized method of characteristics for solving quasilinear
hyperbolic first-order systems can be extended to the case of complex characteristic elements
(see \textit{e.g.} \cite{Perad:1985,Sobolev:1934}). These elements, which had originally been introduced for hyperbolic systems, were generalized to the case of elliptic systems by allowing the wave vectors to be the complex solutions of the dispersion
relation associated with the initial system of equations. We begin by making an algebraization, according to \cite{Burnat:1972,Grundland:1974,Perad:1985}, of
a first-order system of PDEs (\ref{eq:1}) in $p$ independent and $q$ dependent variables written in
the matrix form
\begin{aleq}\label{eq:2.1}
&\mathcal{A}^i(u)u_i=0,\quad i=1,\ldots,p,\\
&x=(x^1,\ldots,x^p)\in X\subseteq \RR^p,\ u=(u^1,\ldots,u^q)\in U\subseteq \RR^q,
\end{aleq}%
where $\mathcal{A}^i(u)=\pa{\mathcal{A}^{\mu i}_\alpha(u)}$ are $m\times q$ matrix functions of $u$ and we denote the partial
derivatives by $u^\alpha_i=\del u^\alpha/\del x^i$. The matrix $L_i^\alpha$ satisfying the
conditions \cite{Burnat:1972}
\begin{eqe}\label{eq:intelem}
u_i^\alpha\in\ac{L_i^\alpha:\mathcal{A}_\alpha^{\mu i}L_i^\alpha=0,\ \mu=1,\ldots,q}
\end{eqe}%
on some open neighborhood of a given point $\tilde{u}\in U$ is called an integral element of the system (\ref{eq:2.1}).
This matrix $L=\pa{{\del u^\alpha}/{\del x^i}}$ is a matrix of the tangent mapping $du\colon X\rarrow
T_uU$ given by the formula
$$X\ni (\delta x^i)\rarrow (\delta u^\alpha)\in T_uU,\quad \text{where }\delta u^\alpha=u_i^\alpha \delta x^i.$$%
The tangent mapping $du(x)$ determines an element of the linear space $L(X,T_uU)$, which can be
identified with the tensor product $T_uU\otimes X^\ast$, (where $X^\ast$ is the dual space of $X$,
\ie the space of linear forms). It is well known \cite{Burnat:1972,Grundland:1974,Perad:1985} that
each element of this tensor product can be represented as a finite sum of simple tensors of the
form
$$L=\gamma\otimes \lambda,$$%
where $\lambda\in X^\ast$ is a covector and $\gamma=\gamma^\alpha\frac{\del}{\del u^\alpha}\in
T_uU$ is a tangent vector at the point $u\in U$. Hence, the integral element $L_i^\alpha$ is called
a simple element if $\rank(L_i^\alpha)=1.$ To determine a simple integral element $L_i^\alpha$ we
have to find a vector field $\gamma\in T_uU$ and a covector $\lambda\in X^\ast$ satisfying the
so-called wave relation
\begin{eqe}\label{eq:2.2}
\pa{\lambda_i\mathcal{A}^{\mu i}_\alpha(u)}\gamma^\alpha=0,\qquad \mu=1,\ldots,m.
\end{eqe}%
The necessary and sufficient condition for the existence of a nonzero solution $\gamma$ for the
equation (\ref{eq:2.2}) is
\begin{eqe}\label{eq:2.3}
\operatorname{rank}\pa{\lambda_i\mathcal{A}^i(u)}<q.
\end{eqe}%
This relation is known as the dispersion relation. If the covector $\lambda=\lambda_i(u)dx^i$
satisfies the dispersion relation (\ref{eq:2.3}) then there exists a polarization vector $\gamma\in
T_uU$ satisfying the wave relation (\ref{eq:2.2}). The algebraic approach which has been used in
\cite{Burnat:1972,Grundland:1974,Perad:1985} for hyperbolic systems of equations (\ref{eq:2.1})
allows the construction of certain classes of $k$-wave solutions admitting $k$ arbitrary functions
of one variable. The replacement of the simple real element $L_i^\alpha$ with a complex element allows us to construct more
general classes of solutions. A specific form of
solution $u(x)$ of the elliptic system (\ref{eq:2.1}) is postulated for which the tangent mapping $du(x)$ is a sum of a
complex element and its complex conjugate
\begin{eqe}\label{eq:2.4}
du^\alpha(x)=\xi(x)\gamma^\alpha(u)\lambda_i(u)dx^i+\bar{\xi}(x)\bar{\gamma}^\alpha(u)\bar{\lambda}_i(u)dx^i,\qquad \alpha=1,\ldots,q,
\end{eqe}%
where $\gamma=(\gamma^1,\ldots, \gamma^q)\in\CC^q$ and $\lambda=(\lambda_1,\ldots,\lambda_p)\in \CC^p$ satisfy
\begin{eqe}\label{eq:star}
\la_i\mathcal{A}_\alpha^{\mu i}(u)\gamma^\alpha=0,\quad \bar{\la}_i\mathcal{A}_\alpha^{\mu i}(u)\bar{\gamma}^\alpha=0.
\end{eqe}%
Here the quantity $\xi(x)\neq 0$ is treated as a complex function of the real variables $x$. In what
follows we assume that the vectors $\gamma$ and $\bar{\gamma}$ are linearly
independent. The proposed form of the solution (\ref{eq:2.4}) is more general than the one proposed in
\cite{Jeffrey:1976,Rozdestvenski:1983} for which the derivatives $u^\alpha_i$ are represented by
a real simple element leading to a simple Riemann wave solution. To distinguish this situation from
the one proposed in (\ref{eq:2.4}), we call the real-valued solution associated with a complex element and its
complex conjugate a simple mode solution in accordance with \cite{Perad:1985}. This means that all first-order derivatives of $u^\alpha$ with respect to $x^i$ are decomposable in the following way
\begin{eqe}\label{eq:star2}
\frac{\del u^\alpha}{\del x^i}=\xi(x)\gamma^\alpha(u)\lambda_i(u)+\bar{\xi}(x)\bar{\gamma}^\alpha(u)\bar{\lambda}_i(u),
\end{eqe}%
where a set of functions $(\lambda,\gamma)$ and their conjugates $(\bar{\lambda},\bar{\gamma})$ on $U$ satisfy the wave relation (\ref{eq:star}). Similarly, as in the case of $k$-waves, for hyperbolic systems \cite{Perad:1985}, we have to find the necessary and sufficient conditions for the existence of solutions of type (\ref{eq:2.4}). These conditions are called involutivity conditions.
\paragraph{}First we derive a number of necessary conditions on the vector fields $(\gamma,\lambda)$ and their
complex conjugate $(\bar{\gamma},\bar{\lambda})$ as a requirement for the existence of rank-$2$
(simple mode) solutions of the homogeneous system (\ref{eq:2.1}). Namely, closing (\ref{eq:2.4}) by
exterior differentiation, we obtain the following 2-forms
\begin{eqe}\label{eq:2.5}
\gamma\otimes(d\xi\wedge \lambda+\xi d\lambda)+\xi d\gamma\wedge
\lambda+\bar{\gamma}\otimes(d\bar{\xi}\wedge
\bar{\lambda}+\bar{\xi}d\bar{\lambda})+\bar{\xi}d\bar{\gamma}\wedge\bar{\lambda}=0,
\end{eqe}%
which have to satisfy (\ref{eq:2.4}). Using (\ref{eq:2.4}) we get
\begin{eqe}\label{eq:2.6}
d\lambda=\bar{\xi}\bar{\lambda}\wedge
\lambda_{,\bar{\gamma}}+\xi\lambda\wedge\bar{\lambda}_{,\gamma},\qquad
d\gamma=\bar{\xi}\gamma_{,\bar{\gamma}}\otimes \bar{\lambda}+\gamma_{,\lambda}\otimes d\lambda,
\end{eqe}%
where we have used the following notation
$$\lambda_{,\gamma}=\gamma^\alpha\frac{\del}{\del u^\alpha}\lambda,\quad \gamma_{,\bar{\gamma}}=\bar{\gamma}^\alpha\frac{\del}{\del u^\alpha}\gamma.$$%
Substituting (\ref{eq:2.6}) into the prolonged system (\ref{eq:2.5}) we obtain
\begin{eqe}\label{eq:2.8}
\gamma\otimes \br{d\xi \wedge \lambda+\xi\bar{\xi}\bar{\lambda}\wedge
\lambda_{,\bar{\gamma}}}+\bar{\gamma}\otimes \br{d\bar{\xi}\wedge
\bar{\lambda}+\xi\bar{\xi}\lambda\wedge\bar{\lambda}_{,\gamma}}+\xi\bar{\xi}\br{\gamma,\bar{\gamma}}\otimes\lambda\wedge\bar{\lambda}=0,
\end{eqe}%
whenever the differential (\ref{eq:2.4}) holds. The commutator of the vector fields $\gamma$ and
$\bar{\gamma}$, is denoted by
$$\br{\gamma,\bar{\gamma}}=\pa{\gamma,\bar{\gamma}}_u+\gamma_{,\lambda_i}\lambda_{i,\bar{\gamma}}-\bar{\gamma}_{,\lambda_i}\lambda_{i,\gamma},$$%
while by $\pa{\gamma,\bar{\gamma}}_u$ we denote a part of the commutator which contains the
differentiation with respect to the variables $u^\alpha$, \ie
$$\pa{\gamma,\bar{\gamma}}_u=\bar{\gamma}^\alpha\frac{\del}{\del u^\alpha}\left.\gamma(u,\lambda)\right|_{\lambda=\mathrm{const}}
-\gamma^\alpha\frac{\del}{\del
u^\alpha}\left.\bar{\gamma}(u,\bar{\lambda})\right|_{\bar{\lambda}=\mathrm{const}}.$$%
Let $\Phi$ be an annihilator of the vectors $\gamma$ and $\bar{\gamma}$, \ie
$$\langle\omega \lrcorner\gamma\rangle=0, \qquad \langle\omega \lrcorner \bar{\gamma}\rangle=0, \quad \omega\in\Phi=\operatorname{An}\ac{\gamma,\bar{\gamma}}.$$%
Here, by the parenthesis $\langle\omega\lrcorner \gamma\rangle$, we denote the contraction of the 1-form
$\omega\in T_u^\ast U$ with the vector field $\gamma\in T_uU$, and we attach a similar meaning to the $\langle\omega\lrcorner\bar{\gamma}\rangle$ for $\bar{\gamma}\in T_uU$. Multiplying the equations (\ref{eq:2.8}) by the 1-form $\omega\in\Phi$ we
get
\begin{eqe}\label{eq:2.9}
\xi\bar{\xi}\langle\omega\lrcorner \br{\gamma,\bar{\gamma}}\rangle\lambda\wedge \bar{\lambda}=0.
\end{eqe}%
We look for the compatibility condition for which the system (\ref{eq:2.9}) does not provide any
algebraic constraints on the coefficients $\xi$ and $\bar{\xi}$. This postulate means that the
profile of the simple modes associated with $\gamma\otimes \lambda$ and $\bar{\gamma}\otimes
\bar{\lambda}$ can be chosen in an arbitrary way for the initial (or boundary) conditions. It
follows from (\ref{eq:2.9}) that the commutator of the vector fields $\gamma$ and $\bar{\gamma}$ is
a linear combination of $\gamma$ and $\bar{\gamma}$, where the coefficients are not necessarily
constant, \ie
%\begin{eqe}\label{eq:2.10}
$$\br{\gamma,\bar{\gamma}}\in\operatorname{span}\ac{\gamma,\bar{\gamma}}.$$%
%\end{eqe}%
So, the Frobenius Theorem is satisfied. That is, at every point $u_0$ of the space of dependent
variables $U$, there exists a tangent surface $S$ spanned by the vector fields $\gamma$ and
$\bar{\gamma}$ passing through the point $u_0\in U$. Moreover the above condition implies that there exists a complex-valued function $\alpha$ for which
\begin{eqe}\label{eq:2.10}
\br{\gamma,\bar{\gamma}}=\alpha\gamma-\bar{\alpha}\bar{\gamma},
\end{eqe}%
since $\overline{\br{\gamma,\bar{\gamma}}}=\br{\bar{\gamma},\gamma}=-\br{\gamma,\bar{\gamma}}$ holds. Next, from the vector fields $\gamma$ and $\bar{\gamma}$ defined on $U$-space, we can
construct the coframe $\Psi$, that is the set of 1-forms $\sigma_1$ and $\sigma_2$ defined on $U$
satisfying the conditions
\begin{aleq}\label{eq:2.11}
&\langle\sigma_1\lrcorner \gamma\rangle=1,\qquad &&\langle\sigma_1\lrcorner\bar{\gamma}\rangle=0,\quad\sigma_1,\sigma_2\in
\Psi,\\
&\langle\sigma_2\lrcorner\gamma\rangle=0, &&\langle\sigma_2\lrcorner \bar{\gamma}\rangle=1.
\end{aleq}%
Substituting (\ref{eq:2.6}) and (\ref{eq:2.10}) into the prolonged system (\ref{eq:2.8}) and multiplying by $\sigma_1$
and $\sigma_2$ respectively, we get
\begin{aleq}\label{eq:2.11b}
&(i)\quad && d\xi\wedge\lambda +\xi\pa{\bar{\xi}\bar{\lambda}\wedge\lambda_{,\bar{\gamma}}+\xi\lambda\wedge\bar{\lambda}_{,\gamma}}+\alpha\xi\bar{\xi}\lambda\wedge\bar{\lambda}=0,\\
&(ii) &&d\bar{\xi}\wedge\bar{\lambda}+\bar{\xi}\pa{\xi\lambda\wedge\bar{\lambda}_{,\gamma}+\bar{\xi}\bar{\lambda}\wedge\lambda_{,\bar{\gamma}}}+\bar{\alpha}\xi\bar{\xi}\bar{\lambda}\wedge\lambda=0.
\end{aleq}%
If $d\xi=d\bar{\xi}$ then equation (\ref{eq:2.11b}.i) is a complex conjugate of (\ref{eq:2.11b}.ii), so one can consider one of them, say (\ref{eq:2.11b}.i). We look for the condition of integrability under which the system (\ref{eq:2.11b}.i) does not impose any restriction on the form of the coefficients $\xi$ and $\bar{\xi}$. Through an exterior multiplication of (\ref{eq:2.11b}.i) by $\lambda$, we obtain
\begin{eqe}\label{eq:2.12i}
\lambda_{,\bar{\gamma}}\wedge \bar{\lambda}\wedge\lambda=0,
\end{eqe}%
and its respective complex conjugate equation
\begin{eqe}\label{eq:2.12ii}
\bar{\lambda}_{,\gamma}\wedge\lambda\wedge\bar{\lambda}=0.
\end{eqe}%
Note that if conditions (\ref{eq:2.10}), (\ref{eq:2.12i}) and (\ref{eq:2.12ii}) are satisfied, then from the Cartan Lemma \cite{Cartan:1953} the system (\ref{eq:2.11b}), and consequently the set of 1-forms (\ref{eq:2.8}), have nontrivial solutions for $d\xi$ and $d\bar{\xi}$. So under these circumstances the following result holds \cite{Perad:1985}.
\begin{prop}\label{prop:1}
The necessary condition for the Pfaffian system (\ref{eq:2.4}) to possess a simple mode solution is that the following two constraints on the complex-valued vector fields $\gamma$ and $\lambda$ be satisfied
$$(i)\ \quad \br{\gamma,\bar{\gamma}}=\alpha \gamma -\bar{\alpha} \bar{\gamma},\quad\text{for any complex-valued function $\alpha\in\CC$,}$$%
$$(ii)\quad \lambda_{,\bar{\gamma}}, \bar{\gamma}_{,\gamma}\in\operatorname{span}\ac{\lambda,\bar{\lambda}}.$$%
\end{prop}%
\paragraph{}Note that conditions (\ref{eq:2.10}), (\ref{eq:2.12i}) and (\ref{eq:2.12ii}) are strong restrictions on the functions $\gamma=\pa{\gamma^1,\ldots,\gamma^q}$ and $\lambda=(\lambda_1,\ldots,\lambda_p)$ and their complex conjugates which satisfy the wave relation (\ref{eq:2.2}). Consequently, the postulated form of a one-mode solution $u(x)$ of the elliptic system (\ref{eq:2.1}), required by the generalized method of characteristics (GMC), is such that all first-order derivatives of $u(x)$ with respect to $x^i$ are decomposable in the form (\ref{eq:2.4}). This restriction is a strong limitation on the admissible class of solutions of (\ref{eq:2.1}). So far, there have been few known examples of such solutions. Therefore, it is worth developing this idea by weakening the integrability conditions (\ref{eq:2.10}), (\ref{eq:2.12i}) and (\ref{eq:2.12ii}) in order to construct multi-mode solutions (considered to be nonlinear superpositions of simple modes). In the next section we propose an alternative way of constructing multi-mode solutions (expressed in terms of Riemann invariants) which are obtained from a version of the conditional symmetry method (CSM) \cite{GrundHuard:2007} by adapting it to the elliptic systems. In Section \ref{sec:5} and \ref{sec:6} a further refinement of the above technique will be presented, which consist of generalizing the idea of mode solutions for quasilinear systems (\ref{eq:2.1}). This idea is based on the specific factorization of integral elements (\ref{eq:intelem}) through the introduction of some rotation matrices by weakening the restrictions imposed by the wave relations (\ref{eq:2.2}). This approach allows us to go deeper into the geometrical aspects of solving the system (\ref{eq:2.1}) and enables us to obtain new results. This constitutes the objective of this paper.
%################################## section 3 ######################################
%\setcounter{equation}{0}%
\section{Conditional symmetry method and multimode solutions in terms of Riemann
Invariants}\label{sec:3}\label{sec:condsym} In this section, we examine certain aspects of the conditional
symmetry method in the context of Riemann invariants for which the wave vectors $\lambda$ and $\bar{\lambda}$ are complex solutions of the dispersion relation (\ref{eq:2.3}) associated with the original system
(\ref{eq:2.1}). Consider a nondegenerate first-order quasilinear elliptic system of PDEs in its matrix form (\ref{eq:2.1}).
Suppose that
there exist $k$ linearly independent non-conjugate complex-valued wave vectors
\begin{eqe}\label{eq:wv}
\la^A(u)=\pa{\la_1^A(u),\ldots,\la_p^A(u)}\in\CC^p,\quad \la^A\neq \bar{\la}^B\quad \forall\
A,B\in\ac{1,\ldots,k},\ k<p,
\end{eqe}%
which satisfy the dispersion relation (\ref{eq:2.3}). One should note that in (\ref{eq:wv}) we do
not require that the indices $A$ and $B$  be distinct, which means that real wave vectors are excluded from our consideration (we do not consider the mixed case for wave vectors involving real and complex
wave vectors). Under the above hypotheses, the $k$ wave vectors (\ref{eq:wv}) and their complex
conjugates
$$\bar{\la}^A(u)=\pa{\bar{\la}_1^A(u),\ldots,\bar{\la}_p^A(u)}\in\CC^p,\quad A=1,\ldots,k,$$%
satisfy the dispersion relation (\ref{eq:2.3}). In what follows it is useful to introduce the
notation c.c. which means the complex conjugate of the previous term or equation. This notation
is convenient for computational purposes allowing the presentation of some expressions in
abbreviated form.
\paragraph{}Suppose that the equation
\begin{eqe}\label{eq:3.5}
u=f(r^1(x,u),\ldots, r^k(x,u),\bar{r}^1(x,u),\ldots,\bar{r}^k(x,u))+\text{c.c.},
\end{eqe}%
implicitly defines a real-valued function $u(x)$ on a neighborhood of the origin $x=0$, which is a solution of the system (\ref{eq:2.1}). Then the complex-valued functions $r^A,\bar{r}^A\colon\RR^p\times \RR^q\rarrow \CC$ are called the
Riemann invariants associated respectively to wave vectors $\lambda^A$, $\bar{\lambda}^A$ and are
defined by
\begin{eqe}\label{eq:3.2}
r^A(x,u)=\lambda_i^A(u)x^i,\quad \bar{r}^A(x,u)=\bar{\lambda}_i^A(u)x^i,\qquad A=1,\ldots,k.
\end{eqe}%
\paragraph{}The Jacobi matrix of derivatives of $u(x)$ is given by
\begin{eqe}\label{eq:3.6}
\del u=(u^\alpha_i)=\pa{I_q-\br{\pa{\frac{\del f}{\del r}+\frac{\del \bar{f}}{\del r}}\frac{\del
r}{\del u}+\text{c.c.}}}^{-1} \pa{\pa{\frac{\del f}{\del r}+\frac{\del\bar{f}}{\del r}}\la+\text{c.c.}},
\end{eqe}%
or equivalently as (see Appendix)
\begin{eqe}\label{eq:3.7}
\del u=\pa{\frac{\del f}{\del r}+\frac{\del \bar{f}}{\del r}}\pa{M^1\lambda+M^2\bar{\lambda}}+\text{c.c.},
\end{eqe}%
where the $k\times k$ matrices $M^1$ and $M^2$ are defined by
\begin{aleq}\label{eq:3.7b}
M^1=&\bigg[I_k-\frac{\del r}{\del u}\pa{\frac{\del f}{\del
r}+\frac{\del\bar{f}}{\del r}}\\
&- \frac{\del r}{\del u}\pa{\frac{\del f}{\del \bar{r}}+\frac{\del\bar{f}}{\del
\bar{r}}}\pa{I_k-\frac{\del\bar{r}}{\del u}\pa{\frac{\del f}{\del \bar{r}}+\frac{\del\bar{f}}{\del
\bar{r}}}}^{-1}\frac{\del\bar{r}}{\del u}\pa{\frac{\del f}{\del
r}+\frac{\del\bar{f}}{\del r}}\bigg]^{-1},\\
M^2&=-M^1 \frac{\del r}{\del u}\pa{\frac{\del f}{\del \bar{r}}+\frac{\del\bar{f}}{\del
\bar{r}}}\pa{I_k-\frac{\del\bar{r}}{\del u}\pa{\frac{\del f}{\del \bar{r}}+\frac{\del\bar{f}}{\del
\bar{r}}}}^{-1},
\end{aleq}%
\begin{eqe}\label{eq:3.8}
\frac{\del f}{\del r}=\pa{\frac{\del f^\alpha}{\del r^A}}\in\CC^{q\times k},\quad \frac{\del
f}{\del \bar{r}}=\pa{\frac{\del f^\alpha}{\del \bar{r}^A}}\in\CC^{q\times k},\quad
\lambda=(\lambda_i^A)\in\CC^{k\times p},
\end{eqe}%
\begin{eqe}\label{eq:3.9}
\frac{\del r}{\del u}=\pa{\frac{\del r^A}{\del u^\alpha}}=\pa{\frac{\del \lambda^A_i}{\del
u^\alpha}x^i}\in\CC^{k\times q},\qquad r=(r^1,\ldots,r^k),
\end{eqe}%
and their respective conjugate equations. The matrices $I_q$ and $I_k$ are the $q\times q$ and
$k\times k$ identity matrices respectively. We use the Implicit Function Theorem to obtain the
following conditions ensuring that functions $r^A,\bar{r}^A$ and $u^\alpha$ are expressible as graphs over
some open subset $\mathcal{D}\subset \RR^p$,
\begin{eqe}\label{eq:3.10}
\det\pa{I_q-\br{\pa{\frac{\del f}{\del r}+\frac{\del \bar{f}}{\del r}}\frac{\del r}{\del
u}+\text{c.c.}}}\neq 0
\end{eqe}%
or
\begin{aleq}\label{eq:3.11}
&\det\pa{I_k-\frac{\del r}{\del u}\pa{\frac{\del f}{\del r}+\frac{\del \bar{f}}{\del r}}}\neq
0,\quad \text{and}\quad\det\Bigg[I_k-\frac{\del \bar{r}}{\del u}\pa{\frac{\del f}{\del
\bar{r}}+\frac{\del \bar{f}}{\del
\bar{r}}} \\
&- \frac{\del \bar{r}}{\del u}\pa{\frac{\del f}{\del r}+\frac{\del \bar{f}}{\del
r}}\pa{I_k-\frac{\del r}{\del u}\pa{\frac{\del f}{\del r}+\frac{\del \bar{f}}{\del
r}}}^{-1}\frac{\del r}{\del u}\pa{\frac{\del f}{\del \bar{r}}+\frac{\del \bar{f}}{\del
\bar{r}}}\Bigg]\neq 0.
\end{aleq}%
So the inverse matrix in (\ref{eq:3.6}) or in (\ref{eq:3.7}) is well-defined in the vicinity of
$x=0$, since
\begin{eqe}\label{eq:3.12}
\frac{\del r}{\del u}=0,\quad\frac{\del\bar{r}}{\del u}=0\ \text{at }x=0.
\end{eqe}%
Note that on the hypersurface defined by equation (\ref{eq:3.5}), where one of the left hand sides of
(\ref{eq:3.10}) or (\ref{eq:3.11}) equal to zero, the gradient of the function $u(x)$ becomes
infinite for some value of $x^i$. So the multimode solution, given by expression (\ref{eq:3.5}),
becomes unbounded on this hypersurface. Consequently, some types of discontinuities, \ie shock waves,
can occur. In what follows, we search for solutions defined on a neighborhood of $x=0$ subject to the conditions (\ref{eq:3.10}) or (\ref{eq:3.11}). This
means that if the initial data is sufficiently small, then there exists a time interval
$\br{t_0,T}$, $T>t_0$ (where we denote the independent variables by $t=x^p, \tilde{x}=(x^1,\dots,
x^n)$ where $n=p-1$) in which the gradient catastrophe for the solution $u(t,\tilde{x})$ of the
system (\ref{eq:2.1}) does not take place \cite{GrundlandVassiliou:1991,Rozdestvenski:1983}.
\paragraph{}Note that the Jacobian matrix of $u(x)$ has at most rank
equal to $2k$. It follows that the proposed solution (\ref{eq:3.5}) is also of rank at most $2k$.
Its image is a $2k$-dimensional submanifold $\Ss_{2k}$ in the space $U$.
\paragraph{}We now introduce a set of $p-2k$ linearly independent
vectors $\xi_a\colon\RR^q\rarrow \CC^p$ defined by
\begin{eqe}\label{eq:3.13}
\xi_a(u)=\pa{\xi_a^1(u),\ldots,\xi_a^p(u)},\quad a=1,\ldots,p-2k,
\end{eqe}%
satisfying the orthogonality conditions
\begin{eqe}\label{eq:3.14}
\lambda_i^A\xi_a^i=0,\qquad \bar{\lambda}_i^A\xi_a^i=0,\quad A=1,\ldots,k,\quad a=1,\ldots,p-2k,
\end{eqe}%
for a set of $2k$ linearly independent wave vectors $\ac{\lambda^1,\ldots,\lambda^k,
\bar{\lambda}^1,\ldots, \bar{\lambda}^k}$. It should be noted that the set
$\ac{\lambda^1,\ldots,\lambda^k,\bar{\lambda^1},\ldots,\bar{\lambda}^k,\xi^1,\ldots,\xi^{p-2k}}$
forms a basis for the space of independent variables $X$. Note also, that the vectors $\xi_a$ are
not uniquely defined since they obey the homogeneous conditions (\ref{eq:3.14}). As a consequence
of equation (\ref{eq:3.5}), the graph $\Gamma=\ac{x,u(x)}$ is invariant under the
family of first-order differential operators
\begin{eqe}\label{eq:3.15i}
X_a=\xi_a^i(u)\frac{\del}{\del x^i},\qquad a=1,\ldots,p-2 k,
\end{eqe}%
defined on $X\times U$ space. Since the vector fields $X_a$ do not include vectors tangent to the
direction of $u$, they form an Abelian distribution on $X\times U$ space, \ie
\begin{eqe}\label{eq:3.16}
\br{X_a,X_b}=0,\qquad a, b=1,\ldots,p-2k.
\end{eqe}%
The set $\ac{r^1,\ldots, r^k,\bar{r}^1,\ldots,\bar{r}^k,u^1,\ldots
u^q}$ constitutes a complete set of invariants of the Abelian algebra $\mathcal{L}$ generated by
the vector fields (\ref{eq:3.15i}). So geometrically, the characterization of the proposed solution
(\ref{eq:3.5}) of the equations (\ref{eq:2.1}) can be interpreted in the following way. If $u(x)$ is a
$q$-component function defined on a neighborhood of the origin $x=0$ such that the graph of the
solution $\Gamma=\ac{(x,u(x))}$ is invariant under a set of $p-2k$ vector fields $X_a$ with the
orthogonality property (\ref{eq:3.14}), then for some function $g$ the expression $u(x)$ is a
solution of equation (\ref{eq:3.5}). Hence the group-invariant solutions of the system
(\ref{eq:2.1}) consist of those functions $u=g(x)$ which satisfy the overdetermined system composed
of the initial system (\ref{eq:2.1}) together with the invariance conditions
\begin{eqe}\label{eq:3.17i}
\xi_a^iu_i^\alpha=0,\qquad i=1,\ldots,p,\quad a=1,\ldots,p-2k,
\end{eqe}%
ensuring that the characteristics of the vector fields $X_a$ are equal to zero.
\paragraph{}It should be noted that, in general, the
conditions (\ref{eq:3.17i}) are weaker than the differential constraints (\ref{eq:2.4}) required by
the generalized method of characteristics, since the latter method is subjected to the algebraic
conditions (\ref{eq:2.2}). In fact, equations (\ref{eq:3.17i}) imply that there exist
complex-valued matrix functions $\Phi_A^\alpha(x,u)$ and $\overline{\Phi}_A^\alpha(x,u)$ defined on
the first jet space $J=J(X\times U)$ such that all first derivatives of $u$ with respect to $x^i$
are decomposable in the following way
\begin{eqe}\label{eq:3.18}
u_i^\alpha=\Phi_A^\alpha(x,u)\la_i^A+\overline{\Phi}_A^\alpha(x,u)\bar{\lambda}_i^A,
\end{eqe}
where
\begin{aleq}\label{eq:3.19}
\Phi_A^\alpha&=\pa{I_q-\br{\pa{\frac{\del f}{\del r}+\frac{\del \bar{f}}{\del r}}\frac{\del r}{\del
u}+\text{c.c.}}}^{-1} \pa{\frac{\del f}{\del
r}+\frac{\del\bar{f}}{\del r}},\\
\overline{\Phi}_A^\alpha&=\pa{I_q-\br{\pa{\frac{\del f}{\del r}+\frac{\del \bar{f}}{\del
r}}\frac{\del r}{\del u}+\text{c.c.}}}^{-1} \pa{\frac{\del f}{\del \bar{r}}+\frac{\del\bar{f}}{\del
\bar{r}}},
\end{aleq}%
or
\begin{aleq}\label{eq:3.20}
\Phi_A^\alpha&=\pa{\frac{\del f}{\del r}+\frac{\del \bar{f}}{\del r}}M^1+\pa{\frac{\del f}{\del \bar{r}}+\frac{\del \bar{f}}{\del \bar{r}}}\overline{M}^2,\\
\overline{\Phi}_A^\alpha&=\pa{\frac{\del f}{\del \bar{r}}+\frac{\del \bar{f}}{\del
\bar{r}}}\overline{M}^1+\pa{\frac{\del f}{\del r}+\frac{\del \bar{f}}{\del r}}M^2.
\end{aleq}%
The matrices $\Phi_A^\alpha\lambda_i^A$ and $\overline{\Phi}_A^\alpha\bar{\lambda}_i^A$ appearing
in equation (\ref{eq:3.18}) do not necessarily satisfy the wave relation (\ref{eq:2.2}). As a
result, the restrictions on the initial data at $t=0$ are eased, so we are able to
consider more diverse types of modes in the superpositions than for the case of the generalized
method of characteristics described in Section \ref{sec:2}.
\paragraph{}We now proceed to solve the overdetermined system
composed of the equations (\ref{eq:2.1}) and the differential constraints (\ref{eq:3.17i})
\begin{eqe}\label{eq:3.21}
\mathcal{A}^{\mu i}_\alpha(u)u_i^\alpha=0,\qquad \xi_a^i(u)u_i^\alpha=0.
\end{eqe}%
Substituting (\ref{eq:3.6}) or (\ref{eq:3.7}) into (\ref{eq:2.1}) yields the trace condition
\begin{eqe}\label{eq:3.22}
\tr\pa{\mathcal{B}^\mu\pa{I_q-\br{\pa{\frac{\del f}{\del r}+\frac{\del \bar{f}}{\del r}}\frac{\del r}{\del
u}+\text{c.c.}}}^{-1} \pa{\pa{\frac{\del f}{\del r}+\frac{\del\bar{f}}{\del r}}\la+\text{c.c.}}}=0,
\end{eqe}%
or
\begin{eqe}\label{eq:3.23}
\tr\pa{\mathcal{B}^\mu\br{\pa{\frac{\del f}{\del r}+\frac{\del \bar{f}}{\del
r}}\pa{M^1\lambda+M^2\bar{\lambda}}+c.c}}=0,
\end{eqe}%
on the wave vectors $\lambda$ and $\bar{\lambda}$ and on the functions $f$ and $\bar{f}$, where
$\mathcal{B}^1,\ldots,\mathcal{B}^m$ are $p\times q$ matrix functions of $u$ (\ie $\mathcal{B}^\mu=(\mathcal{A}^{\mu i}_\alpha(u))\in
\RR^{p\times q}, \mu=1,\ldots, m$). For the given initial system of equations (\ref{eq:2.1}), the
matrices $\mathcal{B}^\mu$ are known functions of $u$ and the trace conditions (\ref{eq:3.22}) or
(\ref{eq:3.23}) are conditions on the functions $f$, $\bar{f}$, $\lambda$, $\bar{\lambda}$ (or on
$\xi$ due to the orthogonality conditions (\ref{eq:3.14})). From the computational point of view,
it is useful to split $x^i$ into $x^{i_p}$ and $x^{i_s}$ and to choose a basis for the wave vectors
$\lambda^A$ and $\bar{\lambda}^A$ such that for each fixed value of the index $A=1,\ldots,k$, we have
\begin{eqe}\label{eq:3.24}
\lambda^A=dx^{i_p}+\lambda_{i_s}^Adx^{i_s},\qquad
\bar{\lambda}^A=dx^{i_p}+\bar{\lambda}_{i_s}^Adx^{i_s},\quad s=1,\ldots,p-1
\end{eqe}%
where $(i_p,i_s)$ is a permutation of $(1,\ldots, p)$. We have just selected a fixed index $i_p$ such that $\lambda_{i_p}\neq 0$ for all $A=1,\ldots,k$, and then we have used the wave vectors $(\lambda^A_{i_p})^{-1}\lambda^A$ instead of the previous $\lambda^A$. So, the expressions (\ref{eq:3.9}) become
\begin{eqe}\label{eq:3.25}
\frac{\del r^A}{\del u^\alpha}=\frac{\del\lambda^A_{i_s}}{\del u^\alpha}x^{i_s},\qquad \frac{\del
\bar{r}^A}{\del u^\alpha}=\frac{\del\bar{\lambda}^A_{i_s}}{\del u^\alpha}x^{i_s}.
\end{eqe}%
Substituting (\ref{eq:3.25}) into (\ref{eq:3.22}) (or (\ref{eq:3.23})) yields
\begin{eqe}\label{eq:3.26}
\tr\pa{\mathcal{B}^\mu\pa{I_q-Q_sx^{i_s}}^{-1}\pa{\frac{\del f}{\del R}+\frac{\del \bar{f}}{\del
R}}\Lambda}=0,\quad \mu=1\ldots,m,
\end{eqe}%
or
\begin{eqe}\label{eq:3.27}
\tr\pa{\mathcal{B}^\mu\pa{\frac{\del f}{\del R}+\frac{\del \bar{f}}{\del
R}}\pa{I_{2k}-K_sx^{i_s}}^{-1}\Lambda}=0,\quad \mu=1\ldots,m,
\end{eqe}%
where $R=(r^1,\ldots, r^k,\bar{r}^1,\ldots,\bar{r}^k)^T$ and
\begin{gaeq}\label{eq:3.28}
Q_s=\pa{\frac{\del f}{\del r}+\frac{\del \bar{f}}{\del r}}\frac{\del \lambda_{i_s}}{\del
u}+\pa{\frac{\del f}{\del \bar{r}}+\frac{\del \bar{f}}{\del \bar{r}}}\frac{\del
\bar{\lambda}_{i_s}}{\del u}=\pa{\frac{\del f}{\del R}+\frac{\del \bar{f}}{\del R}}
\eta_s\in\CC^{q\times q},\\
K_s=\pa{\begin{array}{cc}
          \frac{\del \lambda_{i_s}}{\del u}\pa{\frac{\del f}{\del r}+\frac{\del \bar{f}}{\del r}} & \frac{\del \lambda_{i_s}}{\del u}\pa{\frac{\del f}{\del \bar{r}}+\frac{\del \bar{f}}{\del \bar{r}}} \\
          \frac{\del \bar{\lambda}_{i_s}}{\del u}\pa{\frac{\del f}{\del r}+\frac{\del \bar{f}}{\del r}} & \frac{\del \bar{\lambda}_{i_s}}{\del u}\pa{\frac{\del f}{\del \bar{r}}+\frac{\del \bar{f}}{\del \bar{r}}}
        \end{array}
}=\eta_s\pa{\frac{\del f}{\del R}+\frac{\del \bar{f}}{\del R}}\in\CC^{2k\times 2k}.
\end{gaeq}%
In order to simplify our notation, we denote
\begin{eqe}\label{eq:3.28b}
\Lambda = \pa{\begin{aligned}&\lambda\\
&\bar{\lambda}\end{aligned}}\in\CC^{2k\times p},\qquad \eta_s=\pa{\begin{aligned}&\frac{\del
\lambda_{i_s}}{\del u}\\ &\frac{\del \bar{\lambda}_{i_s}}{\del u}\end{aligned}}\in\CC^{2k\times
q},\qquad \frac{\del f}{\del R}=\pa{\frac{\del f}{\del r}, \frac{\del f}{\del \bar{r}}}\in
\CC^{q\times 2k},
\end{eqe}%
for $i_p$ fixed and $i_s=1,\ldots, p-1$. In (\ref{eq:3.28}) the $2k\times 2k$ matrix $K_s$ is
defined in terms of $k\times k$ subblocks like ${\del \lambda_{i_s}}/{\del u}\pa{{\del
f}/{\del r}+{\del \bar{f}}/{\del r}}$, where $\eta_s$ is the matrix formed of the block
$\del\lambda_{i_s}/\del u$ over the block $\del\bar{\lambda}_{i_s}/\del u$. The notation $\pa{{\del
f}/{\del r}, {\del f}/{\del \bar{r}}}$ represents the matrix formed of the left block $\del f/\del
r$ and the right block $\del f/\del \bar{r}$. Note that the matrix functions $\mathcal{B}^\mu$,
$\del f/\del r$, $\del f/\del \bar{r}$, $\del \bar{f}/\del r$, $\del \bar{f}/\del \bar{r}$, $Q_s$
and $K_s$ depend on $r$ and $\bar{r}$ only. We require that equations (\ref{eq:3.26}) (or (\ref{eq:3.27})) be satisfied for any value of the coordinates $x^{i_s}$. As a consequence, we have some constraints
on these matrix functions.  Thus, using the Cayley-{}-Hamilton Theorem, we can replace equations (\ref{eq:3.26}) by the following condition
\begin{eqe}\label{eq:3.29}
\tr\pa{\mathcal{B}^\mu Q\pa{\frac{\del f}{\del R}+\frac{\del \bar{f}}{\del R}}\Lambda}=0,\qquad\text{where
}Q=\operatorname{adj}(I_q-Q_sx^{i_s})\in\CC^{q\times q},
\end{eqe}%
where $\operatorname{adj}M$ denotes the adjoint of the matrix $M$. As a consequence the matrix $Q$
is a polynomial of order $(q-1)$ in $x^{i_s}$. Taking (\ref{eq:3.29}) and all its partial
derivatives with respect to $x^{i_s}$ (with $r$, $\bar{r}$ fixed at $x=0$), we obtain the following
conditions for the matrix functions $f(r,\bar{r})$ and $\lambda(f(r,\bar{r}))$
\begin{eqe}\label{eq:3.30}
\tr\pa{\mathcal{B}^\mu\pa{\frac{\del f}{\del R}+\frac{\del \bar{f}}{\del R}}\Lambda}=0,\qquad \mu=1,\ldots,m,
\end{eqe}%
\begin{eqe}\label{eq:3.31}
\tr\pa{\mathcal{B}^\mu Q_{\left(s_1\right.}\ldots Q_{s_j)}\pa{\frac{\del f}{\del R}+\frac{\del \bar{f}}{\del
R}}\Lambda}=0,
\end{eqe}%
where $j=1,\ldots,q-1$ and $(s_1,\ldots, s_j)$ denotes the symmetrization over all indices in the
bracket. A similar procedure can be applied to system (\ref{eq:3.27}) to obtain equations (\ref{eq:3.30}) and
\begin{eqe}\label{eq:3.32}
\tr\pa{\mathcal{B}^\mu \pa{\frac{\del f}{\del R}+\frac{\del \bar{f}}{\del R}}K_{(s_1}\ldots
K_{s_j)}\Lambda}=0,
\end{eqe}%
where now $j=1,\ldots,2k-1$. Equations (\ref{eq:3.30}) represent the initial value conditions on a
surface in the space of independent variables $X$, given at $x^{i_s}=0$. Note that equations
(\ref{eq:3.31}) (or (\ref{eq:3.32})) form the conditions required for the preservation of the
property (\ref{eq:3.30}) along the flows represented by the vector fields (\ref{eq:3.15i}). Substituting the expressions (\ref{eq:3.28}) into (\ref{eq:3.31}) or (\ref{eq:3.32}) and simplifying the results, we obtain the unified form
\begin{eqe}\label{eq:3.34}
\tr\pa{\mathcal{B}^{\mu}\pa{\frac{\del f}{\del R}+\frac{\del \bar{f}}{\del
R}}\eta_{\left(s_1\right.}\pa{\frac{\del f}{\del R}+\frac{\del \bar{f}}{\del R}}\ldots
\eta_{{s_j})}\pa{\frac{\del f}{\del R}+\frac{\del \bar{f}}{\del R}}\Lambda}=0.
\end{eqe}%
The index $j$ is either $\max(j)=q-1$ or $\max(j)=2k-1$, we choose the one which is more convenient
from the computational point of view. In this case, for $k\geq 1$ the two approaches, CSM and GMC,
become essentially different and, as we demonstrate in the following example, the CSM can provide
rank-$2k$ solutions which are not Riemann $2k$-waves as defined by the GMC since we weaken the
integrability conditions (\ref{eq:2.1}) for the wave vectors $\lambda$ and $\bar{\lambda}$.
\paragraph{} A change of
variable on $X\times U$ allows us to rectify the vector fields $X_a$ and considerably simplify the
structure of the overdetermined system (\ref{eq:3.21}) which classifies the preceding construction
of multimode solutions. For this system, in the new coordinates, we derive the necessary and
sufficient conditions for the existence of rank-$2k$ solutions of the form (\ref{eq:3.5}). Suppose
that there exists an invertible $2k\times 2k$ subblock matrix
\begin{eqe}\label{eq:3.35}
H=(\Lambda^s_t), \quad 1\leq s,t\leq 2k,
\end{eqe}%
of the larger matrix $\Lambda\in \CC^{2k\times p}$, then the independent vector fields $X_a$, given by (\ref{eq:3.15i}), can be
written as
\begin{eqe}\label{eq:3.36}
\tilde{Z}_a=\del_{x^{a+2k}}-(H^{-1})^B_t\Lambda^t_{a+2k}\del_{x^B},\quad B=1,\ldots,2k,
\end{eqe}%
which have the required form (\ref{eq:3.15i}) and for which the orthogonality conditions
(\ref{eq:3.14}) are fulfilled. We introduce new coordinate functions
\begin{gaeq}\label{eq:3.37}
z^1=r^1(x,u),\ldots, z^{k}=r^{k}(x,u),z^{k+1}=\bar{r}^1(x,u),\ldots, z^{2k}=\bar{r}^{k}(x,u),\\
z^{2k+1}=x^{2k+1},\ldots,z^p=x^p,\quad v^1=u^1,\ldots, v^q=u^q,
\end{gaeq}%
on $X\times U$ space which allow us to rectify the vector fields (\ref{eq:3.36}). As a result, we
get
\begin{eqe}\label{eq:3.38}
Z_1=\frac{\del}{\del z^{2k+1}},\quad \ldots,\quad Z_{p-2k}=\frac{\del}{\del z^p}.
\end{eqe}%
The $p$-dimensional submanifold invariant under $Z_{1},\ldots,Z_{p-2k}$, is defined by equations of
the form
\begin{eqe}\label{eq:3.39}
v=f(z^1,\ldots, z^k,\bar{z}^1,\ldots, \bar{z}^k)+\text{c.c.}
\end{eqe}%
for an arbitrary function $f\colon X\rarrow U$. The expression (\ref{eq:3.39}) is the general solution of
the invariance conditions
\begin{eqe}\label{eq:3.40}
v_{z^{2k+1}}=0,\quad\ldots, \quad v_{z^p}=0.
\end{eqe}%
In general, the initial system (\ref{eq:2.1}) described in the new coordinates $(x,v)\in X\times U$
is a nonlinear system of first-order PDEs, of the form
\begin{eqe}\label{eq:3.42}
\begin{aligned}
&\mathcal{A}^{l\mu}_\beta(v)\frac{\del r^i}{\del x^l}\frac{\del v^\beta}{\del z^j}=0,\quad
&&i=1,\ldots, k,\\
&\mathcal{A}^{l\mu}_\beta(v)\frac{\del \bar{r}^i}{\del x^l}\frac{\del v^\beta}{\del z^j}=0,\quad &&i=k+1,\ldots, 2k,\\
&\mathcal{A}^{i\mu}_\beta(v)\frac{\del v^\beta}{\del z^j}=0, &&i=2k+1,\ldots,p.
\end{aligned}
\end{eqe}%
We obtain the following Jacobi matrix in the coordinates $(z,\bar{z},v)$
\begin{eqe}\label{eq:3.43}
\frac{\del z^j}{\del x^i}=\pa{\Omega^{-1}}^j_l\Lambda^l_i\in \CC^{p\times p},\qquad
\Omega=\delta_s^i-\frac{\del \Lambda^i_l}{\del v^\beta}\frac{\del v^\beta}{\del z^s}x^l,
\end{eqe}%
whenever the invariance conditions (\ref{eq:3.40}) are satisfied. Appending to the system
(\ref{eq:3.42}) the invariance condition (\ref{eq:3.40}), we obtain the quasilinear reduced system
of PDEs
\begin{gaeq}\label{eq:3.44}
\tr\pa{\mathcal{B}^\mu(v)\pa{I_{q}-\frac{\del v}{\del z}\frac{\del R}{\del
v}}^{-1}\frac{\del v}{\del z}\Lambda}=0,\qquad \mu=1,\ldots,m,\\
 \frac{\del v}{\del z^{2k+1}}=0,\quad \ldots,\quad \frac{\del v}{\del z^p}=0,
\end{gaeq}%
or
\begin{gaeq}\label{eq:3.45}
\tr\pa{\mathcal{B}^\mu(v)\frac{\del v}{\del z}\pa{I_{2k}-\frac{\del R}{\del
v}\frac{\del v}{\del z}}^{-1}\Lambda}=0,\qquad \mu=1,\ldots,m,\\
\frac{\del v}{\del z^{2k+1}}=0,\quad \ldots,\quad\frac{\del v}{\del z^p}=0.
\end{gaeq}%
\paragraph{}We now provide some basic definitions that are required in order to be able to encompass Riemann invariants into the conditional symmetry method.
\paragraph{}A vector field $Z_a$ is called a conditional symmetry of
the original system (\ref{eq:2.1}) if $Z_a$ is tangent to the manifold $\Ss=\Ss_\Delta\cap \Ss_Q$,
\ie
\begin{eqe}\label{eq:3.46}
\left.\pr^{(1)}Z_a\right|_\Ss\in T_{(x,u^{(1)})}\Ss,
\end{eqe}%
where the first prolongation of $Z_a$ is given by
\begin{eqe}\label{eq:3.47}
\pr^{(1)}Z_a=Z_a-\xi^i_{a,u^\beta}u_j^\beta u_i^\alpha\frac{\del}{\del u^\alpha_j},\qquad
a=1,\ldots,p-2k
\end{eqe}%
and the submanifolds of the solution spaces are given by
\begin{eqe}\label{eq:3.48}
\Ss_\Delta=\ac{(x,u^{(1)}):\mathcal{A}_\alpha^{i\mu}u_i^\alpha=0,\qquad \mu=1,\ldots,m},
\end{eqe}%
and
\begin{eqe}\label{eq:3.49}
\Ss_Q=\ac{(x,u^{(1)})\colon\xi_a^i(u)u_i^\alpha=0,\quad \alpha=1,\ldots, q,\ a=1,\ldots,p-2k}.
\end{eqe}%
Consequently, an Abelian Lie algebra $\mathcal{L}$ generated by the vector
fields $Z_1,\ldots, Z_{p-2k}$ is called a conditional symmetry algebra of the original system
(\ref{eq:2.1}) if the conditions
\begin{eqe}\label{eq:3.50}
\left.\pr^{(1)}Z_a\pa{\mathcal{A}^iu_i}\right|_\Ss=0,\quad a=1,\ldots, p-2k,
\end{eqe}
are satisfied.
\paragraph{}Supposing that $\mathcal{L}$, spanned by the vector fields $Z_1,\ldots,Z_{p-2k}$, is a conditional symmetry algebra of the system (\ref{eq:2.1}), a solution $u=g(x)$ is
said to be a conditionally invariant solution of the system (\ref{eq:2.1}) if the graph
$\Gamma=\ac{(x,g(x))}$ is invariant under the vector fields $Z_1,\ldots, Z_{p-2k}$.
\paragraph{\textbf{Proposition} 2.} \textit{
A nondegenerate quasilinear first-order elliptic system of PDEs (\ref{eq:2.1}) in $p$ independent
variables and $q$ dependent variables admits a ($p-2k$)-dimensional conditional symmetry algebra $\mathcal{L}$ if and only if $p-2k$ linearly independent vector fields $Z_1,\ldots, Z_{p-2k}$ satisfy the
conditions (\ref{eq:3.30}) and (\ref{eq:3.34}) on some neighborhood of $(x_0,u_0)$ of $\Ss$. The
solutions of (\ref{eq:2.1}) which are invariant under the Abelian Lie algebra $\mathcal{L}$ are precisely
$k$-mode solutions of the form (\ref{eq:3.5}). }
\paragraph{Proof.}The proof of this proposition is essentially similar to that of the proposition in
\cite{GrundHuard:2007}. We express the vector
fields $Z_a$ in the new coordinates $(z,v)$ on $X\times U$. Equations (\ref{eq:3.38}) and
(\ref{eq:3.47}) imply that
\begin{eqe}\label{eq:3.60}
\pr^{(1)}Z_a=Z_a,\quad a=1,\ldots, p-2k.
\end{eqe}%
The symmetry criterion that has to be satisfied for $G$ to be the symmetry group of the
overdetermined system (\ref{eq:3.44}) (or (\ref{eq:3.45})) requires that the vector fields $X_a$ of
$G$ satisfy $Z_a(\Delta)=0$, whenever equation (\ref{eq:3.44}) (or (\ref{eq:3.45})) is satisfied.
Thus the symmetry criterion applied to the invariance conditions (\ref{eq:3.40}) vanishes
identically. Applying this criterion to the system (\ref{eq:3.42}) in the new coordinates, carrying
out the differentiation and taking into account the conditions (\ref{eq:3.30}) and (\ref{eq:3.34}),
we obtain that the equations are identically satisfied.
\paragraph{}The converse is also true. The assumption that the
system (\ref{eq:2.1}) is nondegenerate means, according to \cite{Olver:Application_of_Lie}, that it
is locally solvable and is of maximal rank at every point $(x_0,u_0)\in S$. Therefore, the
infinitesimal symmetry condition is a necessary and sufficient condition for the existence of the
symmetry group $G$ of the overdetermined system (\ref{eq:3.21}). Since the vector fields $Z_a$ form
an Abelian distribution on $X\times U$, it follows that conditions (\ref{eq:3.30}) and
(\ref{eq:3.34}) are satisfied. The solutions of the overdetermined system (\ref{eq:3.21}) are
invariant under the algebra $\mathcal{L}$ generated by the $p-2k$ vector fields
$Z_1,\ldots,Z_{p-2k}$. The invariants of the group $G$ of such vector fields are provided by the
functions $\ac{r^1,\ldots, r^k,\bar{r}^1,\ldots, \bar{r}^k,u^1,\ldots,u^q}$. So the general
multimode solution of (\ref{eq:2.1}) takes the required form (\ref{eq:3.5}).\hspace*{1.4in}$\Box$
\paragraph{} It should be noted that the decomposition of the Jacobian matrix (\ref{eq:3.18}) is less restrictive than the one proposed by the generalized method of characteristics (\ref{eq:star2}). The trace conditions (\ref{eq:3.30}) and (\ref{eq:3.34}) are weaker than the compatibility conditions (\ref{eq:2.10}), (\ref{eq:2.12i}) and (\ref{eq:2.12ii}) required by the generalized method of characteristics. This analysis has led us to consider a new approach which weakens the decomposition structure of the solution in terms of the wave matrix $\Lambda$ and the polarisation matrix $(\del f/\del r+\del \bar{f}/\del r)$ determined by the equations (\ref{eq:3.30}) and (\ref{eq:3.34}). The new proposed geometrical approach based on the introduction of rotational matrices, allows us to find new results presented in sections IV and V.
%################

%########
%########################################## section 4 ##############################################
%\setcounter{equation}{0}%
%############ nonhom#########################
\section{Simple wave solutions of an inhomogeneous quasilinear system}\label{sec:ms}\label{sec:5}
Consider an inhomogeneous first-order system of $q$ quasilinear PDEs in $p$ independent variables and $q$ unknowns of the form
\begin{eqe}\label{eq:SW:1}
\mathcal{A}^{\alpha i}_\beta(u) u^\beta_i=b^\alpha(u),\qquad \alpha,\beta=1,\ldots,q,\quad i=1,\ldots,p.
\end{eqe}%
Let us underline that the system can be either hyperbolic or elliptic. We are looking for real solutions describing the propagation of a simple wave which can be realized by the system (\ref{eq:SW:1}). We postulate a form of the solution $u$ in terms of a Riemann invariant $r$, \ie
\begin{eqe}\label{eq:SW:2}
u=f(r),\qquad r=\lambda_i(u) x^i, \quad i=1,\ldots,p,
\end{eqe}%
where $\lambda(u)=\pa{\lambda_1(u),\ldots,\lambda_p(u)}$ is a real-valued wave vector.
We evaluate the Jacobian matrix $u^\beta_i$ by applying the chain rule:
$$u^\beta_i=\frac{\del f^\beta}{\del r}\pa{r_{x^i}+r_{u^\alpha}u^\alpha_i}=\frac{\del f^\beta}{\del r}\pa{\lambda_i+\lambda_{i,u^\alpha}x^iu^\alpha_i}.$$%
We assume that the matrix
\begin{eqe}\label{eq:SW:4}
\Phi=\pa{I_q-\frac{d f}{d r}\frac{\del r}{\del u}}\in \RR^{q\times q},
\end{eqe}%
is invertible, where we have used the following notation $d f/d r=\pa{d f^1/d r,\ldots, d f^q/d r}^T$, $\del r/\del u =\pa{\del r/\del u^1,\ldots,\del r/\del u^q}$. The Jacobian matrix $\del u$ takes the form
\begin{eqe}\label{eq:SW:5}
\del u=\Phi^{-1}\frac{d f}{d r}\lambda\in\RR^{q\times q}.
\end{eqe}%
Replacing the Jacobian matrix (\ref{eq:SW:5}) into the original system (\ref{eq:SW:1}), we get
\begin{eqe}\label{eq:SW:7}
\mathcal{A}^{\alpha i}_\beta\pa{\Phi^{-1}}^\beta_\mu\frac{d f^\mu}{d r}\lambda_i=b^\alpha.
\end{eqe}%
Note that the expression $\Phi^{-1}{d f}/{d r}\in \RR^q$ is a contravariant vector as well as $b\in\RR^q$. Hence there exists a nonzero scalar function $\Omega=\Omega(x,u)$, a rotation matrix $L=L(x,u)\in SO(q)$ and a vector $\tau=\tau(x,u)\in\RR^q$ such that
\begin{eqe}\label{eq:SW:8}
\Phi^{-1}\frac{d f}{d r}=\Omega L b+\tau,\qquad \mathcal{A}^i\lambda_i\tau=0.
\end{eqe}%
It should be noted that we assume appropriate levels of differentiability of the functions $\Omega$, $\lambda$, $L$, $\tau$ and $b$, as necessary in order to justify all the following steps. Using relation (\ref{eq:SW:8}), we eliminate the vector $\Phi^{-1}{d f}/{d r}\in \RR^q$ from equation (\ref{eq:SW:7}), which allows us to factor out the vector $b$ on the right after regrouping all terms on the left of the resulting equation. Therefore, we obtain the condition
\begin{eqe}\label{eq:SW:13}
\pa{\Omega\mathcal{A}^{i}\lambda_i L-I_q}b=0,
\end{eqe}%
on the scalar function $\Omega$, the wave vector $\lambda$ and the rotation matrix $L$. This implies that we have the following dispersion relation
\begin{eqe}\label{eq:SW:14}
\det\pa{\Omega\mathcal{A}^{i}\lambda_i L-I_q}=0.
\end{eqe}%
Once a scalar function $\Omega$, a wave vector $\lambda$ and a matrix $L$ satisfying (\ref{eq:SW:13}) have been obtained, equation (\ref{eq:SW:8}) must be used in order to determine the function $f$. Replacing the expression (\ref{eq:SW:4}) for the matrix $\Phi$ into equation (\ref{eq:SW:8}) and solving for the vector ${d f}/{d r}$ and taking into account the relation $\del r/\del u=(\del \lambda_i/\del u) x^i$, we find that
\begin{eqe}\label{eq:SW:15}
\frac{d f}{d r}=\frac{\Omega L b+\tau}{1+(\del \lambda_i/\del u) (\Omega L b+\tau)x^i},
\end{eqe}%
which cannot admit the gradient catastrophe. Indeed, if we suppose that $1+(\del \lambda_i/\del u)(\Omega L b+\tau) x^i=0$ when we proceed from equation (\ref{eq:SW:8}) to equation (\ref{eq:SW:15}), then we conclude that $(\Omega L b+\tau)=0$. Consequently, since the matrix $\Phi$ is invertible, therefore we conclude from (\ref{eq:SW:8}) that the solution for $f(r)$ is constant, so it cannot admit the gradient catastrophe.
\paragraph{}Up till now, we cannot be sure that the system (\ref{eq:SW:15}) for $f(r)$ is well-defined in the sense that it represents a system for $f(r)$ expressed in terms of $r$ only. To ensure this, we begin by introducing vector fields orthogonal to the wave vector $\lambda$, that is, vector fields of the form
\begin{eqe}\label{eq:SW:16}
X_a=\xi^i_a(u)\del_{x^i},\qquad a=1,\ldots,p-1,\quad i=1,\ldots,p,
\end{eqe}%
where
\begin{eqe}\label{eq:SW:17}
\xi_a^i\lambda_i=0,\qquad a=1,\ldots,p-1,\quad\rank(\xi_a^i)=p-1.
\end{eqe}%
Consequently, the wave vector $\lambda$ together with the vectors $\xi_a$, $a=1,\ldots,p-1$ form a basis for $\RR^p$. Moreover, let us note that the vector fields $X_a$ form an Abelian Lie algebra of dimension $p-1$. Since we suppose that $f=f(r)$, the vector fields (\ref{eq:SW:16}) annihilate the left hand side of the equation (\ref{eq:SW:15}), then it have to cancels out the right side of the equation (\ref{eq:SW:15}). Therefore, the conditions for the system of ODEs (\ref{eq:SW:15}) to be well-defined in terms of $r$ are
\begin{eqe}\label{eq:SW:18}
X_a\br{\frac{\Omega L b+\tau}{1+ \frac{\del \lambda_i}{\del u}(\Omega L b+\tau)x^i}}=0,\qquad a=1,\ldots,p-1.
\end{eqe}%
\paragraph{}In summary, system (\ref{eq:SW:1}) admits a simple wave solution if the following conditions are satisfied:
\begin{itemize}
\item[  i)] there exist a scalar function $\Omega(x,u)$, a wave vector $\lambda(u)$ and a rotation matrix $L(x,u)$ satisfying the wave relation (\ref{eq:SW:13});%
\item[ ii)] there exist locally $p-1$ vector fields (\ref{eq:SW:16})  which satisfy the orthogonality relation (\ref{eq:SW:17});%
\item[iii)] the right-hand side of equation (\ref{eq:SW:15}) is annihilated by the vector fields (\ref{eq:SW:16}), \ie conditions (\ref{eq:SW:18}) are satisfied;%
\item[ iv)] $\det \Phi\neq 0$, where $\Phi$ is given by (\ref{eq:SW:4});%
\item[  v)] $1+ \frac{\del \lambda_i}{\del u}(\Omega Lb+\tau)x^i\neq 0$, $\lambda\in C^1$.
\end{itemize}%
These conditions are sufficient, but not necessary, since the vanishing of $\det\Phi$ does not imply that solutions of form (\ref{eq:SW:2}) do not exist. We finish the present analysis of the simple wave solutions of system (\ref{eq:SW:1}) with several remarks:
\begin{itemize}
\item[1)] In general, there are more parameters (arbitrary functions) defining $\Omega$ and the rotation matrix $L$ than the minimum number required to satisfy condition (\ref{eq:SW:13}). The remaining arbitrary quantities are arbitrary functions of the $x$'s and $u$'s, which have to be used to satisfy the conditions (\ref{eq:SW:18}). However, in the particular case when the system (\ref{eq:SW:1}) has two equations in two dependent variables, the two-dimensional matrix $L$ is defined by a single parameter and $\Omega$ is the only other parameter available to satisfy the condition (\ref{eq:SW:13}). Thus, since the system is autonomous (it can be expressed solely in terms of the $u$'s and their derivatives), it is clear that, in that case, $\Omega$ and $L$ depend only on the dependent variables $u$'s.
\item[2)] Supposing that $b$ is continuous and assuming that conditions (i)-{}-(v) are satisfied, the right side of equation (\ref{eq:SW:15}) is continuous, which ensures the existence and uniqueness of the solution for $f(r)$ (see for example \cite{Ince}).%
\item[3)] A sufficient condition for the system (\ref{eq:SW:15}) to be expressible in terms of $r$ only (\ie well-defined) is:
$$\frac{\del \lambda_i}{\del u}L b=0,\qquad i=1,\ldots, p.$$%
This condition is trivially satisfied when the vector $\lambda$ is constant. More generally, it is sufficient to require that $\frac{\del \lambda_i}{\del u}Lbx^i$ be proportional to $r=\lambda_i x^i$, \ie that there exist $\beta(x,u)$ such that
$$\frac{\del \lambda_i}{\del u}L b=\beta(x,u)\lambda_i,\qquad i=1,\ldots,p.$$%
\item[4)] If the matrix $A^i\lambda_i$ is invertible, then
$$\frac{\del \lambda_i}{\del u}Lbx^i=\frac{\del \lambda_i}{\del u}(\mathrm{A}^j\lambda_j)^{-1}(\mathrm{A}^j\lambda_j) Lb x^i=\frac{1}{\Omega}\frac{\del \lambda_i}{\del u}(\mathcal{A}^j\lambda_j)^{-1}bx^i$$%
is satisfied due to condition (\ref{eq:SW:13}). From the previous remark, we deduce
$$\frac{1}{\Omega}\frac{\del \lambda_i}{\del u}(\mathcal{A}^ j\lambda_j)^{-1}b=\beta(u) \lambda_i,\quad i=1,\ldots,p,$$%
which is a sufficient condition if the rotation matrix $L$ is not involved when the matrix $A^j\lambda_j$ is invertible.
\end{itemize}%
We note that the approach presented above generalizes the results obtained in \cite{GrundlandZelazny:1983} where the simple states have been constructed with wave vectors $\lambda$ of constant direction for hyperbolic inhomogeneous systems (\ref{eq:SW:1}).
%############## simple mode #########################
\section{Simple mode solutions for an inhomogeneous quasilinear system}\label{sec:6}
We now generalize the concept of simple wave solutions for inhomogeneous, quasilinear, systems of form (\ref{eq:SW:1}).  As in the case of the simple wave, the system can be either hyperbolic or elliptic. We look for a real solution, in terms of a Riemann invariant $r$ and its complex conjugate $\bar{r}$, of the form
\begin{eqe}\label{eq:ms:1}
u=f(r,\bar{r}),\qquad r(u,x)=\lambda_i(u)x^i,\quad \bar{r}(u,x)=\bar{\lambda}_i(u)x^i,\quad i=1,\ldots, p,
\end{eqe}%
where $\lambda(u)=\pa{\lambda_1(u),\ldots,\lambda_p(u)}$ is a complex wave vector and $\bar{\lambda}(u)$ is its complex conjugate. The Jacobian matrix takes the form
\begin{eqe}\label{eq:ms:2}
\del u=\Phi^{-1}\pa{\frac{\del f}{\del r}\lambda +\frac{\del f}{\del \bar{r}} \bar{\lambda}}\in \RR^{q\times p},
\end{eqe}%
where we assume that the matrix
\begin{eqe}\label{eq:ms:3}
\Phi=\pa{I_q-\frac{\del f}{\del r}\frac{\del r}{\del u}-\frac{\del f}{\del\bar{r}}\frac{\del\bar{r}}{\del u}}\in\RR^{q\times q}.
\end{eqe}%
is invertible. Replacing the Jacobian matrix (\ref{eq:ms:2}) into the system (\ref{eq:SW:1}), we obtain
\begin{eqe}\label{eq:ms:4}
\mathcal{A}^{i}\Phi^{-1}\pa{\frac{\del f}{\del r}\lambda_i +\frac{\del f}{\del \bar{r}} \bar{\lambda}_i}=b,\qquad \mathcal{A}^i=\pa{A^{\mu i}_\alpha}\in\RR^{q\times q}.
\end{eqe}%
We introduce a rotation matrix $L=L(x,u)\in SO(q,\CC)$ and its complex conjuguate $\bar{L}=\bar{L}(x,u)\in SO(q,\CC)$ for which the relations
\begin{eqe}\label{eq:ms:5}
\Phi^{-1}\frac{\del f}{\del r}=\Omega L b + \tau,\qquad \Phi^{-1}\frac{\del f}{\del \bar{r}}=\bar{\Omega} \bar{L} b+ \bar{\tau},
\end{eqe}%
hold, where $\Omega(x,u)$ and its conjugate $\bar{\Omega}$ are scalar complex functions, while $\tau(x,u)$ and its complex conjugate vector $\bar{\tau}(x,u)$ satisfy
\begin{eqe}\label{eq:ms:7}
\mathcal{A}^i\lambda_i \tau+\mathcal{A}^i\bar{\lambda}_i \bar{\tau}=0.
\end{eqe}%
The vectors $\tau$ and $\bar{\tau}$ can be seen as characteristic vectors of the homogeneous part of equation (\ref{eq:ms:4}). We eliminate the vectors $\Phi^{-1}(\del f/\del r)$ and $\Phi^{-1}(\del f/\del \bar{r})$ from equation (\ref{eq:ms:4}) using the equations (\ref{eq:ms:5}). Considering the equation (\ref{eq:ms:7}), we obtain, as a condition on functions $\Omega$, $\lambda$, $L$ and their complex conjugates, the relation
\begin{eqe}\label{eq:ms:8}
\pa{\mathcal{A}^i\pa{\lambda_i\Omega L+\bar{\lambda}_i\bar{\Omega}\bar{L}}-I_q}b=0.
\end{eqe}%
The vector $b$ can be identified with an eigenvector of equation (\ref{eq:ms:8}). Since $b$ is known from the initial system (\ref{eq:SW:1}), the equation has to be satisfied by an appropriate choice of  functions $\Omega$, $\bar{\Omega}$, of wave vectors $\lambda$, $\bar{\lambda}$ and of rotation matrices $L$ and $\bar{L}$. In particular, this requires that the dispersion relation
\begin{eqe}\label{eq:ms:10}
\det\pa{\mathcal{A}^i\pa{\lambda_i L+\bar{\lambda}_i\bar{L}}-I_q}=0
\end{eqe}%
hold. Equation (\ref{eq:ms:10}) is an additional condition on $\lambda$ and $\bar{\lambda}$, $L$ and $\bar{L}$ for the equation (\ref{eq:ms:8}) to have a solution. Multiplying equations (\ref{eq:ms:5}) on the left by the matrix $\Phi$, writing the matrix $\Phi$ explicitly using the notations (\ref{eq:ms:3}), and then solving for $\del f/\del  r$ and $\del f/\del\bar{r}$, we obtain the system
\begin{aleq}\label{eq:ms:12}
\frac{\del f}{\del r}=&\frac{(1+\bar{\sigma_1})(\Omega L b+\tau)-\sigma_2 (\bar{\Omega}\bar{L}b+\bar{\tau})}{|1+\sigma_1|^2-|\sigma_2|^2},\\
\frac{\del f}{\del \bar{r}}=&\frac{(1+\sigma_1)(\bar{\Omega}\bar{L}b+\bar{\tau})-\bar{\sigma}_2(\Omega L b+\tau)}{|1+\sigma_1|^2-|\sigma_2|^2},
\end{aleq}%
where the scalar functions $\sigma_1$ and $\sigma_2$  and their complex conjugates are defined by the equations
\begin{eqe}\label{eq:ms:13}
\sigma_1=\frac{\del r}{\del u}(\Omega L b+\tau),\qquad \sigma_2=\frac{\del \bar{r}}{\del u}(\Omega L b+\tau).
\end{eqe}%
In order to ensure that system (\ref{eq:ms:12}) is well-defined in the sense that it can be expressed as a system for $f$ in terms of $r$ and $\bar{r}$, we introduce the vector fields
\begin{eqe}\label{eq:ms:14}
X_a=\xi^i_a(u)\del_{x^i},\qquad a=1,\ldots, p-2,
\end{eqe}%
where the complex coefficients $\xi^i_a(u)$ satisfy the orthogonality relations
\begin{eqe}\label{eq:ms:15}
\xi^i_a\lambda_i=0,\qquad \xi^i_a\bar{\lambda}_i=0.
\end{eqe}%
Next, we apply the vector fields (\ref{eq:ms:14}) to equations (\ref{eq:ms:12}), which gives us the following conditions
\begin{aleq}
&X_a\br{\frac{(1+\bar{\sigma_1})(\Omega L b+\tau)-\sigma_2 (\bar{\Omega}\bar{L}b+\bar{\tau})}{|1+\sigma_1|^2-|\sigma_2|^2}}=0,\\
&X_a\br{\frac{(1+\sigma_1)(\bar{\Omega}\bar{L}b+\bar{\tau})-\bar{\sigma}_2(\Omega L b+\tau)}{|1+\sigma_1|^2-|\sigma_2|^2}}=0,
\end{aleq}%
$a=1,\ldots, p-2$, since the vector fields $X_a$ annihilate all the functions $f(r,\bar{r})$.
\paragraph{Remark 1:}The system (\ref{eq:ms:12}) for $f(r,\bar{r})$ is not necessarily integrable. However, it is possible to use the arbitrary functions defining the function $\Omega(x,u)$, the wave vectors $\lambda(u)$, $\bar{\lambda}(u)$, the vectors $\tau(x,u)$, $\bar{\tau}(x,u)$ and the matrices of rotation $L(x,u)$, $\bar{L}(x,u)$, in order to satisfy the compatibility conditions of the system (\ref{eq:ms:12}).
%######## exemple de Liouville ########
\section{Simple examples}\label{sec:7}
We give some simple examples to illustrate the construction introduced in Sections \ref{sec:5} and \ref{sec:6}.
\subsection{Simple wave solution of the inhomogeneous Gibbon-{}-Tsarev system}
Consider the inhomogeneous hydrodynamic-type system for $q=2$ unknown functions, $u$ and $v$, and $p=2$ independent variables, $t$ and $x$, of form
\begin{eqe}\label{eq:GT:1}
\mathcal{A}^1\frac{\del}{\del t}\pa{\begin{array}{c}
                                      u \\
                                      v
                                    \end{array}
}+\mathcal{A}^2\frac{\del}{\del x}\pa{\begin{array}{c}
                                        u \\
                                        v
                                      \end{array}
}=\pa{\begin{array}{c}
        b_1 \\
        b_2
      \end{array}
},
\end{eqe}%
where
\begin{gaeq}\label{eq:GT:2}
\mathcal{A}^1=\pa{\begin{array}{cc}
                    1 & 0 \\
                    0 & 1
                  \end{array}
},\quad \mathcal{A}^2=\pa{\begin{array}{cc}
                            -v & 0 \\
                            0 & -u
                          \end{array}
},\\
b=\pa{b_1,b_2}^{T}=(u-v)^{-1}\pa{-1,1}^T.
\end{gaeq}%
We look for a simple wave solution ($k=1$). Therefore, in accordance with the method presented in Section \ref{sec:5}, we have to find a scalar function $\Omega(x,u)$, a rotation matrix $L(x,u)\in SO(2,\RR)$,
a wave vector $\lambda(u)\in\RR^2$ and a characteristic vector $\tau(x,u)$ which satisfy the algebraic equations (\ref{eq:SW:8}) and (\ref{eq:SW:13}). It is easily verified that equation (\ref{eq:SW:13}) has no solution when equation (\ref{eq:SW:8}) is solved in such a way that $\tau\neq0$. Hence, for the rest of this example, we consider the case when $\tau=0$. The functions $\Omega$ and $L$ which satisfy condition (\ref{eq:SW:13}) take the form
\begin{gaeq}\label{eq:GT:3}
L=\pa{\begin{array}{cc}
        \cos\theta & -\sin\theta \\
        \sin\theta & \cos\theta
      \end{array}
},\qquad \theta=\arctan\pa{\frac{-(u-v)\lambda_2}{2\lambda_1-(u-v)\lambda_2}},\\
\Omega=\frac{2}{\lambda_2 (u-v)\sin\theta+(2\lambda_1-(u+v)\lambda_2)\cos\theta},
\end{gaeq}%
where the components $\lambda_1$ and $\lambda_2$ of the wave vector $\lambda$ are arbitrary functions of $u$ and $v$. Since $\tau=0$, the reduced system (\ref{eq:GT:3}) takes the form
\begin{eqe}\label{eq:GT:4}
\frac{df}{dr}=\frac{\Omega L b}{1+\Omega\pa{\frac{\del\lambda_i}{\del u}}L b x^i}.
\end{eqe}%
Since the solution (\ref{eq:GT:3}) for $\Omega$ and $L$ depends only on $u$ and $v$, the system is well-defined in terms of the Riemann invariant $r$ only if the scalar function
\begin{eqe}\label{eq:GT:5}
\pa{\frac{\del \lambda_i}{\del u}}Lbx^i
\end{eqe}%
depends on $r$ only. In order to ensure this requirement, we introduce a vector field of form (\ref{eq:SW:16}) satisfying the orthogonality condition (\ref{eq:SW:17}). This vector field takes the explicit form
\begin{eqe}\label{eq:GT:6}
X_r=-\lambda_2 \del_t+\lambda_1\del_x.
\end{eqe}%
We apply the vector field field (\ref{eq:GT:6}) to the expression (\ref{eq:GT:5}) and we require that the result be zero. This leads us to the PDE for the components $\lambda_1$ and $\lambda_2$ of the wave vector in terms of the variables $u$ and $v$
\begin{eqe}\label{eq:GT:7}
 (\lambda_1-u\lambda_2)\pa{\lambda_2\frac{\del \lambda_1}{\del v}-\lambda_1\frac{\del \lambda_2}{\del u}}+ (\lambda_1-v\lambda_2)\pa{-\lambda_2\frac{\del \lambda_1}{\del v}+\lambda_1 \frac{\del \lambda_1}{\del v}}=0
\end{eqe}%
If $\lambda_1$ and $\lambda_2$ satisfy the PDE (\ref{eq:GT:7}), then the expression (\ref{eq:GT:5}) can be written
\begin{eqe}\label{eq:GT:8}
\frac{(\lambda_1-u\lambda_2)\pa{-\lambda_1\frac{\del \lambda_1}{\del u}-\lambda_2\frac{\del \lambda_2}{\del u}}+(\lambda_1-v\lambda_2)\pa{\lambda_1\frac{\del \lambda_1}{\del v}+\lambda_2 \frac{\del \lambda_2}{\del v}}}{(\lambda_1^2+\lambda_2^2)(u-v)}r
\end{eqe}%
In order to simplify the integration procedure of system (\ref{eq:GT:4}), we require that the coefficient in front of $r$ in expression (\ref{eq:GT:8}) cancel out. This requirement leads to an autonomous system. Therefore, we have to solve the system consisting of the PDE (\ref{eq:GT:7}) and the following PDE
\begin{eqe}\label{eq:GT:9}
(\lambda_1-u\lambda_2)\pa{-\lambda_1\frac{\del \lambda_1}{\del u}-\lambda_2\frac{\del \lambda_2}{\del u}}+(\lambda_1-v\lambda_2)\pa{\lambda_1\frac{\del \lambda_1}{\del v}+\lambda_2 \frac{\del \lambda_2}{\del v}}=0.
\end{eqe}%
Taking the linear combination of equations (\ref{eq:GT:7}) and (\ref{eq:GT:9}) of form $\lambda_1$ (\ref{eq:GT:7}) $+$ $\lambda_2$ (\ref{eq:GT:9}) and $\lambda_2$ (\ref{eq:GT:7}) $-$ $\lambda_1$ (\ref{eq:GT:9}), we find respectively the PDEs
\begin{eqe}\label{eq:GT:10}
\lambda_1\pa{\frac{\del \lambda_2}{\del u}-\frac{\del \lambda_2}{\del v}}-\lambda_2\pa{u\frac{\del \lambda_2}{\del u}-v\frac{\del \lambda_2}{\del v}}=0,
\end{eqe}
\begin{eqe}\label{eq:GT:11}
\lambda_1\pa{\frac{\del \lambda_1}{\del u}-\frac{\del \lambda_1}{\del v}}-\lambda_2\pa{u\frac{\del\lambda_1}{\del u}-v\frac{\del\lambda_1}{\del v}}=0.
\end{eqe}%
From equation (\ref{eq:GT:11}), we find
\begin{eqe}\label{eq:GT:12}
\lambda_2=\pa{u\frac{\del \lambda_1}{\del u}-v\frac{\del \lambda_1}{\del v}}^{-1}\pa{\frac{\del \lambda_1}{\del u}-\frac{\del \lambda_1}{\del v}}\lambda_1.
\end{eqe}%
Substituting $\lambda_2$ given by (\ref{eq:GT:12}) into the PDE (\ref{eq:GT:10}), we find the second-order PDE
\begin{aleq}\label{eq:GT:13}
&4\,\eta\, \left( {\frac {\partial \lambda_{{1}}}{\partial \xi}}
   \right) ^{2}{\frac {\partial ^{2}\lambda_{{1}}}{\partial
{\eta}^{2}}} +4\,\eta\,
 \left( {\frac {\partial \lambda_{{1}} }{\partial \eta}} \right) ^{2}{\frac {\partial ^{2}\lambda_{{1}}}{\partial {\xi}^{2}}}
-8\,\eta\, \left( {
\frac {\partial \lambda_{{1}}}{\partial \xi}}
 \right)  \left( {\frac {\partial \lambda_{{1}}}{\partial \eta}}  \right) {\frac {\partial ^{2} \lambda_{{1}}}{\partial \xi
\partial \eta}}\\
 \quad &+4\, \left( {\frac {\partial \lambda_{{1}}}{
\partial \eta}}  \right) ^{3}-4\,
 \left( {\frac {\partial \lambda_{{1}}}{\partial \xi}} \right) ^{2}{\frac {\partial \lambda_{{1}}}{\partial \eta}}=0,
\end{aleq}%
where $\xi=u+v$, $\eta=u-v$. Through a separation of variables of the form $\lambda_1(\xi,\eta)=p(\xi)q(\eta)$, we obtain the solution
\begin{eqe}\label{eq:GT:14}
p(\xi)=\alpha_1\exp(\alpha_0 \xi),\qquad q(\eta)=\exp\pa{\alpha_2^{-1}(1+\alpha_2^2\eta^2)^{1/2}+\alpha_3},
\end{eqe}%
where $\alpha_0,\ldots,\alpha_3$ are integration constants. The corresponding solution for $\lambda_1$ expressed in terms of the variables $u$ and $v$ takes the form
\begin{eqe}\label{eq:GT:15}
\lambda_1=\alpha_1\exp\pa{\alpha_0\alpha_2^{-1}\Theta},\quad \Theta=\alpha_2(u+v)+\epsilon(1+\alpha_2^2(u-v)^2)^{1/2}.
\end{eqe}%
Introducing $\lambda_1$ given by (\ref{eq:GT:15}) into equation (\ref{eq:GT:12}), we obtain
$$
\lambda_2=2 \frac{\epsilon_1\alpha_1\alpha_2}{\Theta}\exp\pa{\alpha_0\alpha_2^{-1}\Theta}.
$$%
In order to simplify the integration procedure for the reduced system (\ref{eq:GT:4}), we let $\alpha_0=0$ and $\alpha_1=1$. Therefore, the components of the wave vectors reduce to
\begin{eqe}\label{eq:GT:16}
\lambda_1=1,\qquad\lambda_2=\frac{2 \alpha_2}{\Theta}.
\end{eqe}%
Replacing the functions $\Omega$ and $L$ given by (\ref{eq:GT:3}), where $\lambda_1$ and $\lambda_2$ are defined by (\ref{eq:GT:16}), into the system (\ref{eq:GT:4}) leads to the system
\begin{aleq}\label{eq:GT:17}
\frac{du}{dr}=&-\frac{\alpha_2(u+v)+\epsilon(1+\alpha_2^2(u-v)^2)^{1/2}}{(u-v) (\alpha_2(u-v)+(1+\alpha_2^2(u-v)^2)^{1/2})},\\
\frac{dv}{dr}=&\frac{\alpha_2(u+v)+\epsilon(1+\alpha_2^2(u-v)^2)^{1/2}}{(u-v) (-\alpha_2(u-v)+(1+\alpha_2^2(u-v)^2)^{1/2})}.
\end{aleq}%
The solution of the system (\ref{eq:GT:17}) takes the implicit form
\begin{aleq}\label{eq:GT:18}
&u=\frac{1}{2}\beta_1+\frac{1}{2\alpha_2^2(2v-\beta_1)},\\
&2 \alpha_2 v+\frac{1}{2v-\beta_1}=4 \alpha_2^4 \beta_1 (r+\beta_3),
\end{aleq}%
where $\beta_1$, $\beta_2$ and $\beta_3$ are integration constants. Introducing the solution for $u$, given by (\ref{eq:GT:18}), we obtain that the wave vector (\ref{eq:GT:15}) becomes $\lambda=(\lambda_1,\lambda_2)=(1,2\beta_1^{-1})$. Hence, $r=t+2\beta_1^{-1}x$, and the explicit solution of (\ref{eq:GT:1}) takes the form
\begin{aleq}
u(t,x)=&\frac{1}{2}\beta_1+\frac{1}{2\alpha_2^2(2v(t,x)-\beta_1)},\\
v(t,x)=&{\frac {{\alpha_{{2}}}^{2}\beta_{{1}}+\eta+\sqrt {{\alpha_{{2}}}^
{4}{\beta_{{1}}}^{2}-2\,{\alpha_{{2}}}^{2}\beta_{{1}}\eta+{\eta}^{2}-4
\,{\alpha_{{2}}}^{2}}}{4{\alpha_{{2}}}^{2}}},
\end{aleq}%
where
$$\eta=4\alpha_2^4\beta_1(r+\beta_3)=4\alpha_2^4\beta_1(t+2\beta_1^{-1} x+\beta_3).$$%

\subsection{Nonlinear interaction of waves and particles}
Consider the inhomogeneous system describing the propagation
of shock waves in the nonlinear interaction of waves and
particles \cite{Luneburg:1964}
\begin{eqe}\label{eq:e1:1}
u_x+\phi_y=2^{1/2}a \exp(u/2)\sin(\phi/2),\quad u_y-\phi_x=-2^{1/2}a\exp(u/2)\cos(\phi/2).
\end{eqe}%
It should be noted that the compatibility condition of the mixed derivatives of $\phi$ corresponds to the Liouville equation
\begin{eqe}\label{eq:e1:2}
u_{xx}+u_{yy}=a^2\exp{u}.
\end{eqe}%
Therefore, each solution of the system (\ref{eq:e1:1}) also gives us a solution of the Liouville equation (\ref{eq:e1:2}). We use the methods presented in Section \ref{sec:6} to obtain the general solution of the system (\ref{eq:e1:1}). First, write the system (\ref{eq:e1:1}) in the matrix form
\begin{eqe}\label{eq:e1:3}
\pa{\begin{array}{c c}
1 & 0\\
0 & -1
\end{array}}\pa{\begin{array}{c }
u_x\\
\phi_x
\end{array}}+\pa{\begin{array}{c c}
0 & 1\\
1 & 0
\end{array}}\pa{\begin{array}{c}
u_y\\
\phi_y\end{array}}=\pa{\begin{array}{c}
b_1\\
b_2\end{array}},
\end{eqe}%
where
$$b_1=2^{1/2}a \exp(u/2)\sin(\phi/2),\qquad b_2=2^{1/2}a\exp{u/2}\cos(\phi/2).$$%
The matrices $\mathcal{A}^i$ introduced in equation (\ref{eq:ms:4}) are given by
$$\mathcal{A}^1=\pa{\begin{array}{c c}
1 & 0\\
0 & -1
\end{array}},\qquad \mathcal{A}^2=\pa{\begin{array}{c c}
0 & 1\\
1 & 0
\end{array}}.$$%
Condition (\ref{eq:ms:8}) is then satisfied by the scalar function $\Omega$, the wave vector $\lambda$ and the rotation matrix $L$, defined by
$$\Omega=12^{1/4}(1-\epsilon^i),\quad \lambda=(1,i),\quad L=\pa{\begin{array}{c c} l_{11} & l_{12}\\
l_{21} & l_{22}\end{array}},\quad \epsilon=\pm 1,$$%
together with their complex conjugates, where
$$\begin{aligned}
l_{11}&=l_{22}=-108^{1/4}i\pa{\frac{3^{1/2}\epsilon (b_1+ib_2)^2+i(b_1-ib_2)^2}{6(1-\epsilon i) (b_1^2+b_2^2)}}\\
l_{12}&=-l_{21}=108^{1/4}i\pa{\frac{3^{1/2}\epsilon i (b_1+ib_2)^2+(b_1-ib_2)^2}{6(1-\epsilon i) (b_1^2+b_2^2)}}
\end{aligned}$$%
Since the wave vectors $\lambda$ and $\bar{\lambda}$ are constant, the quantities $\sigma_1$ and $\sigma_2$ (and their complex conjugates) vanish. So, we obtain the system (\ref{eq:ms:12}) written in terms of the Riemann invariants $r=x+ i y$ and $\bar{r}=x- i y$
\begin{aleq}\label{eq:e1:5}
u_r&=24^{-1/2}a\pa{i(3^{1/2}+3 i \epsilon)\cos(\phi/2)+(3^{1/2}-3 i \epsilon)\sin(\phi/2)}\exp(u/2)+\sigma\\
\phi_r&=24^{-1/2}a\pa{(3^{1/2}-3 i \epsilon)\cos(\phi/2)-i(3^{1/2}+3 i \epsilon)\sin(\phi/2)}\exp(u/2)+i\sigma\\
u_{\bar{r}}&=24^{-1/2}a\pa{-i(3^{1/2}-3 i \epsilon)\cos(\phi/2)+(3^{1/2}+3 i \epsilon)\sin(\phi/2)}\exp(u/2)+\bar{\sigma}\\
\phi_{\bar{r}}&=24^{-1/2}a\pa{(3^{1/2}-3 i \epsilon)\cos(\phi/2)+i(3^{1/2}-3 i \epsilon)\sin(\phi/2)}\exp(u/2)-i\bar{\sigma}\\
\end{aleq}%
where $\sigma(r,\bar{r})$ is an arbitrary scalar function defining the vector $\tau=(\sigma,i \sigma)^T$, which satisfies condition (\ref{eq:ms:7}). The system (\ref{eq:e1:5}) can be written in the more compact form
\begin{aleq}\label{eq:e1:6}
&a)\quad &&\frac{\del }{\del r}\pa{u+i \phi}=2^{-1/2}a i \exp\pa{\frac{u-i\phi}{2}},\\
&b)\quad &&\frac{\del }{\del \bar{r}}\pa{u-i \phi}=-2^{-1/2}a i \exp\pa{\frac{u+i\phi}{2}},\\
&c)\quad &&\frac{\del }{\del r}\pa{u-i \phi}=-(3/2)^{1/2}a \exp\pa{\frac{u+i\phi}{2}}+2\sigma,\\
&d)\quad &&\frac{\del }{\del \bar{r}}\pa{u+i \phi}=-(3/2)^{1/2}a \exp\pa{\frac{u-i\phi}{2}}+2\bar{\sigma}.\\
\end{aleq}%
Since the equations (\ref{eq:e1:6}.c) and (\ref{eq:e1:6}.d) are complex conjugates they can be satisfied by choosing the quantity $\sigma$ which annihilates (\ref{eq:e1:6}.c). Consequently, only the two equations (\ref{eq:e1:6}.a) and (\ref{eq:e1:6}.b) remain to solve. After the change of variable
\begin{eqe}\label{eq:e1:7}
f=\exp\pa{\frac{u+i \phi}{2}},
\end{eqe}%
and $\bar{f}$ is the complex conjugate of $f$. Equations (\ref{eq:e1:6}.a) and (\ref{eq:e1:6}.b) take the form
\begin{eqe}\label{eq:e1:8}
f_r=2^{-3/2} i a|f|^2,\qquad \bar{f}_{\bar{r}}=-2^{-3/2} i a|f|^2,
\end{eqe}%
which have general solution
\begin{eqe}\label{eq:e1:8}
f(r,\bar{r})=\frac{-8^{1/2}i\bar{\psi'}(\bar{r})}{a (\psi(r)+\bar{\psi}(\bar{r}))},
\end{eqe}%
where $\psi$ is an arbitrary function of $r$ and $\bar{\psi}$ its complex conjugate. Replacing the solution (\ref{eq:e1:8}) into equations (\ref{eq:e1:7}), solving for $u$ and $\phi$, and using the Riemann invariants $r$ and $\bar{r}$, we obtain the general solution of the system (\ref{eq:ms:1}). This solution depends on one arbitrary complex function $\psi$ of one complex variable $r$ and its complex conjugate function.
\begin{aleq}
u&=2\ln\pa{\frac{8|\psi'(x+iy)|}{a^2(\psi(x+i y)+\psi(x-i y))^2}},\\
\phi&=-i\ln\pa{-\frac{\bar{\psi}'(x-i y)}{\psi'(x + i y)}}+2 n\pi,
\end{aleq}%
where $n$ is odd and the term $2 n\pi$ is associated with the admissible branches of the logarithm. It is easily verified that the solution for $u$ satisfies the Liouville equation (\ref{eq:e1:2}).
\subsection{The ideal plastic flow}\label{sec:4}
In this section we would like to illustrate the proposed approach for constructing simple mode
solutions with the example of an ideal nonstationary irrotational planar plastic flow
subjected to an external force due to a work function $V$ (potential if $V_t=0$). Under the above assumptions the examined model is governed by a quasilinear elliptic homogenous system of five equations in (2+1) dimensions of the
form \cite{Chak:2006,Hill:1998,Kat:1}
\begin{aleq}\label{eq:4.1}
&(\mathrm{a})\ &&\s_x- \pa{\q_x \cos 2\q+\q_y \sin2\q}+\p \pa{V_x-u_t - u u_x - v u_y}=0,\\
&(\mathrm{b}) &&\s_y- \pa{\q_x\sin2\q - \q_y \cos 2\q}+\p \pa{V_y-v_t - u v_x - v v_y}=0,\\
&(\mathrm{c}) &&(u_y+v_x)\sin 2\q + (u_x-v_y)\cos2\q=0,\\
&(\mathrm{d}) && u_x+v_y=0.
\end{aleq}%
The independent variables are denoted by $(x^i)=(t,x,y)\in X\subset \RR^3$
and the unknown functions by $(u^\alpha)=(\sigma,\theta, u,v)\in U\subset \RR^4$. The potential $V$ is a given function of $(t,x,y)\in\RR^3$. The stress tensor is defined
by the mean pressure $\sigma$ and the angle $\theta$ made by the first principal components of the stress tensor relative to the $x$-axis minus $\pi/4$. Equation (\ref{eq:4.1}.c) represents the Saint-Venant--von Mises plasticity equation. Equation
(\ref{eq:4.1}.d) for the velocities $u$ and $v$ (along the $x$-axis and
$y$-axis respectively) corresponds to the incompressibility of
the plastic material flow. The first-order
partial derivatives of $\sigma$ with respect to the independent variables $x$ and $y$ are denoted
$\sigma_x$ and $\sigma_y$ respectively. The first-order partial derivatives of $u$, $v$ and
$\theta$ are denoted similarly.
\paragraph{}Before using the approach described in Section \ref{sec:6} to obtain
solutions of the system (\ref{eq:4.1}), we begin by expressing the system (\ref{eq:4.1}) in a form
which is more convenient for the purpose of computation. In the first place, we note that the
equations (\ref{eq:4.1}.a) and (\ref{eq:4.1}.b) governing the average pressure $\sigma$ can be
solved by quadrature when the compatibility condition for the mixed
derivatives of $\sigma$ with respect to $x$ and $y$ is satisfied. This equation takes the form
\begin{aleq}\label{eq:4.2}
&2\pa{\theta_{x}^2-\theta_{xy}-\theta_y^2}\cos(2\theta)+\pa{\theta_{xx}-4\theta_x\theta_y-\theta_{yy}}\sin(2\theta)\\
&\ -\rho\pa{(u_y-v_x)_t+u(u_y-v_x)_x+v(u_y-v_x)_y}=0.
\end{aleq}%
If the equation (\ref{eq:4.2}) holds, then the pressure $\sigma$ can be expressed in terms of the
velocities $u$, $v$, of the angle $\theta$ and of the potential $V$ with the form
\begin{aleq}\label{eq:4.3}
\sigma(t,x,y)&=-\rho V(t,x,y)+{1\over 2}\sin(2\theta(t,x,y))+\rho
\frac{u(t,x,y)^2+v(t,x,y)^2}{2}\\&\ +\rho\int u_t(t,x,y)dx
+\int \theta_y(t,x,y)\sin(2\theta(t,x,y))dx+c_0(t).
\end{aleq}%
The problem of obtaining a solution of system (\ref{eq:4.1}) can be reduced to that of solving the system consisting of equations (\ref{eq:4.1}.c), (\ref{eq:4.1}.d) and (\ref{eq:4.2}). Since equation (\ref{eq:4.2}) is neither quasilinear nor first-order, the approach developed in Section \ref{sec:6} cannot be used. Nevertheless, this approach can be used for the system (\ref{eq:4.1}.c) and (\ref{eq:4.1}.d) in order to obtain a solution for the components of the velocity, assuming that $\theta$ is a real-valued function that can be expressed in terms of a Riemann invariant $r=\lambda_1(u,v)x+\lambda_2(u,v)y$ and its complex conjugate. In matrix form, the system consisting of (\ref{eq:4.1}.c) and (\ref{eq:4.1}.d) is of the form
\begin{eqe}\label{eq:p:1}
\mathcal{A}^1\frac{\del}{\del x}\pa{\begin{array}{c}
                                       u \\
                                       v
                                     \end{array}
}+\mathcal{A}^2\frac{\del}{\del_y}\pa{\begin{array}{c}
                                       u \\
                                       v
                                     \end{array}}=0,
\end{eqe}%
where the $2\times 2$ matrices $\mathcal{A}^1$ and $\mathcal{A}^2$ are of the form
\begin{eqe}\label{eq:p:2}
\mathcal{A}^1=\pa{\begin{array}{cc}
                    \cos(2\theta) & \sin(2\theta) \\
                    1 & 0
                  \end{array}
},\quad \mathcal{A}^2=\pa{\begin{array}{cc}
                    \sin(2\theta) & -\cos(2\theta) \\
                    0 & 1
                  \end{array}}.
\end{eqe}%
Since the system (\ref{eq:p:1}) is homogeneous, the condition (\ref{eq:ms:8}) is identically satisfied since $b=0$. We have to ensure that condition (\ref{eq:ms:7}) is satisfied by an appropriate choice of wave vector $\lambda$, of the characteristic vector $\tau$ and of their respective complex conjugates. Substituting the matrices $\mathcal{A}^i$, given by (\ref{eq:p:2}), into equation (\ref{eq:ms:7}), we obtain the conditions
\begin{gaeq}\label{eq:p:3}
\pa{t_1-\mu_1 t_2+\mu_2 T_2}\cos(2\theta)+\pa{\mu_1 t_1-\mu_2 T_1+t_2}\sin(2\theta)=0\\
t_1+\mu_1 t_2-\mu_2 T_2=0.
\end{gaeq}%
where $\lambda_1=1$, $\lambda_2=\mu_1+i \mu_2$, $\pa{\tau=t_1+i T_1, t_2+iT_2}$. The real and imaginary parts of $\lambda_2$ are functions of $u$ and $v$, while the components of $\tau$ are functions of $x,y,u$ and $v$. In order to simplify our calculations, we assume for the rest of this example that $\mu_1=0$, $\mu_2$ is a real constant and $\tau$ is a function of $r$ and $\bar{r}$ only. Under the above assumptions, the coefficients $\sigma_1$ and $\sigma_2$ in the system (\ref{eq:ms:12}) vanish. Therefore, the system becomes
\begin{eqe}\label{eq:p:4}
\frac{\del}{\del r}\pa{\begin{array}{c}
                         u \\
                         v
                       \end{array}
}=\tau,\quad \frac{\del}{\del \bar{r}}\pa{\begin{array}{c}
                         u \\
                         v
                       \end{array}
}=\bar{\tau}.
\end{eqe}%
Solving the equations (\ref{eq:p:3}) for $\tau_1$ and $\tau_2$, and then replacing the obtained result into the system (\ref{eq:p:4}), we obtain the following system
\begin{aleq}\label{eq:p:5}
\frac{\del u}{\del r}=&iT_1+\mu_2 T_2,\\
\frac{\del v}{\del r}=&\mu_2T_1+\pa{i-\frac{2\mu_2\cos(2\theta)}{\sin(2\theta)}}T_2,\\
\frac{\del u}{\del\bar{r}}=&-iT_1+\mu_2T_2,\\
\frac{\del v}{\del\bar{r}}=&\mu_2T_1+\pa{-\frac{2\mu_2\cos(2\theta)}{\sin(2\theta)}-i}T_2.
\end{aleq}%
Since the functions $T_1$ and $T_2$ are arbitrary, we can define them by the following equations
$$T_1=-\frac{i}{2}\pa{\frac{\del u}{\del r}-\frac{\del u}{\del \bar{r}}},\quad T_2=-\frac{i}{2}\pa{\frac{\del v}{\del r}-\frac{\del v}{\del\bar{r}}}.$$%
As a result of this choice, the system (\ref{eq:p:5}) reduces to the following two PDEs
\begin{eqe}\label{eq:p:7}
\frac{\del(u+iv)}{\del r}=-\frac{\del (u-iv)}{\del\bar{r}},\\
\end{eqe}
\begin{eqe}\label{eq:p:8}
\frac{\del(u-iv)}{\del r}-2\frac{\cos(2\theta)}{\sin(2\theta)}\pa{\frac{\del v}{\del r}-\frac{\del v}{\del \bar{r}}}=\frac{\del(u+iv)}{\del\bar{r}}.
\end{eqe}%
A trivial solution of equation (\ref{eq:p:7}) is given by
$$u+iv=2\bar{h}(\bar{r}),\qquad u-iv=2h(r),$$%
where $h(r)$ is an arbitrary function of the Riemann invariant $r=x+iy$ and $\bar{h}$ is its complex conjugate. The solution for the components of the velocity, $u$ and $v$, takes the form
\begin{eqe}\label{eq:4.13}
u=h(r)+\bar{h}(\bar{r}),\qquad v=i\pa{h(r)-\bar{h}(\bar{r})}.
\end{eqe}%
From equation (\ref{eq:p:8}), we find that the function $\theta$ takes the form
\begin{eqe}\label{eq:4.14}
\theta=\frac{1}{2}\arctan\pa{i\frac{h^{(1)}(r)+\bar{h}^{(1)}(\bar{r})}{\pa{h^{(1)}(r)-\bar{h}^{(1)}(\bar{r})}}},
\end{eqe}%
where we denote $h^{(n)}=d^nh/dr^n$. Next, we substitute (\ref{eq:4.14}) into equation
(\ref{eq:4.2}) in order to determine the function $h$ and its complex conjugate. Proceeding in this
way, we find that $h$ and $\bar{h}$ satisfy the third-order ODE which is separable in $r$ and
$\bar{r}$. It is therefore equivalent to the system consisting of the equation
\begin{eqe}\label{eq:4.15}
-2\frac{h^{(3)}(r)}{h^{(1)}(r)^2}+3\frac{h^{(2)}(r)^2}{h^{(1)}(r)^3}=\Omega,
\end{eqe}%
and its respective complex conjugate equation, where $\Omega$ is a real separation constant. Defining
\begin{eqe}\label{eq:4.16}
g(r)=h^{(1)}(r),
\end{eqe}
equation (\ref{eq:4.15}) can be written as
\begin{eqe}\label{eq:4.17}
2gg''-3 (g')^2+\Omega g^3=0.
\end{eqe}%
Equation (\ref{eq:4.17}) admits a first integral
%\begin{eqe}\label{eq:4.18}
%g^{3/2} g'+\Omega \ln g=c_1
%\end{eqe}%
which can be rewritten in the equivalent form
\begin{eqe}\label{eq:4.19}
(g^{-3/2} \Omega^{-1/2}(c_1- \ln g)^{-1/2}) dg=dr.
\end{eqe}%
By the change of variable
\begin{eqe}\label{eq:4.20}
g=\exp\pa{\frac{c_1-\omega^2}{\Omega}},
\end{eqe}%
equation (\ref{eq:4.19}) can be transformed to $2\exp\pa{(2\Omega)^{-1}(\omega^2-c_1)}d\omega=\pm
\Omega dr,$ for which the solution, in term of the inverse error function
$\operatorname{erf}^{-1}$, is
\begin{eqe}\label{eq:4.21}
\omega=\sqrt{-2\Omega\br{\operatorname{erf}^{-1}\pa{e^{c_1/(2\Omega)}\frac{-i(c_2\pm\Omega
r)}{\sqrt{2\pi \Omega}}}}^2},
\end{eqe}%
where $c_2$ is a constant of integration. We substitute (\ref{eq:4.20}) into (\ref{eq:4.16}) with
$\omega$ given by (\ref{eq:4.21}) and integrate the result. We obtain the function $h(r)$ and its
complex conjugate in terms of the error functions $\operatorname{erfi}$ and
$\operatorname{erf}^{-1}$ as
\begin{aleq}\label{eq:4.22}
h(r)&=-\frac{2\pi
c_1}{\Omega}\operatorname{erfi}\br{\operatorname{erf}^{-1}\pa{c_2+c_1r}}+c_3,\\
\bar{h}(\bar{r})&=-\frac{2\pi
\bar{c}_1}{\Omega}\operatorname{erfi}\br{\operatorname{erf}^{-1}\pa{\bar{c}_2+\bar{c}_1\bar{r}}}+\bar{c}_3,
\end{aleq}%
where the $c_i\in\CC$ are integration constants and the real separation constant $\Omega$ is non
zero. Since the expressions (\ref{eq:4.22}) are solutions of the system (\ref{eq:4.15}), we have that $\theta$ given
by (\ref{eq:4.14}) satisfies the compatibility condition (\ref{eq:4.2}) for $\sigma$. One should
note that no derivatives with respect to time appear in the PDE (\ref{eq:4.2}). Consequently, the
equation (\ref{eq:4.2}) is still satisfied if, in equations (\ref{eq:4.13}) and (\ref{eq:4.14}), we
replace the function $h(r)$ and its complex conjugate respectively by
\begin{aleq}\label{eq:4.22bis}
h(t,r)&=-\frac{2\pi
c_1(t)}{\Omega(t)}\operatorname{erfi}\br{\operatorname{erf}^{-1}\pa{c_2(t)+c_1(t)r}}+c_3(t),\\
\bar{h}(t,\bar{r})&=-\frac{2\pi
\bar{c}_1(t)}{\Omega(t)}\operatorname{erfi}\br{\operatorname{erf}^{-1}\pa{\bar{c}_2(t)+\bar{c}_1(t)\bar{r}}}+\bar{c}_3(t),
\end{aleq}%
which are obtained by substituting the arbitrary complex functions $c_i(t)$ in place of the
integration constant $c_i$ and the arbitrary real function $\Omega(t)$ in place of the separation
constant $\Omega$. Hence, the general solution of the system (\ref{eq:4.1}) takes the form
\begin{aleq}\label{eq:4.23}
u(t,x,y)&=h(t,r)+\bar{h}(t,\bar{r}),\\
v(t,x,y)&=i\pa{h(t,r)-\bar{h}(t,\bar{r})},\\
\theta(t,x,y)&=\frac{1}{2}\arctan\pa{i\frac{h_r(t,r)+\bar{h}_r(t,\bar{r})}{\pa{h_r(t,r)-\bar{h}_r(t,\bar{r})}}},\\
\sigma(t,x,y)&=-\rho V(t,x,y)+{1\over 2}\sin(2\theta(t,x,y)) +\rho
\frac{u(t,x,y)^2+v(t,x,y)^2}{2}\\ &\ +\rho\int u_t(t,x,y)dx
+\int \theta_y(t,x,y)\sin(2\theta(t,x,y))dx+\sigma_0(t),
\end{aleq}%
where the functions $h$ and $\bar{h}$ are defined by (\ref{eq:4.22bis}). Note that the
solution (\ref{eq:4.23}) is a real-valued solution.
\par
\begin{figure}
\begin{center}
\includegraphics[width=3.5in]{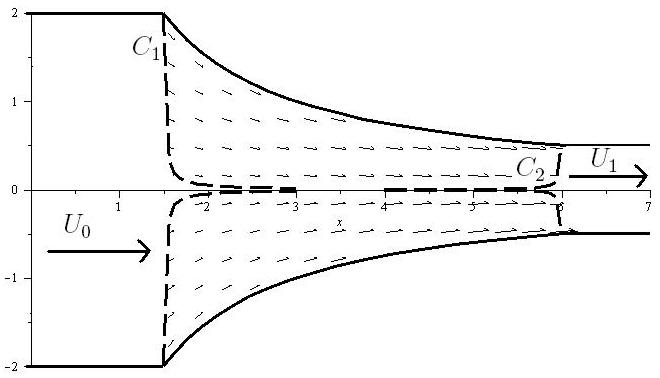}
\end{center}
\caption{Extrusion die corresponding to the solution (\ref{eq:4.25}).} \label{fig:1}
\end{figure}

 Consider the situation when
$\Omega$ is identically equal to zero in the system of ODEs (\ref{eq:4.15}). This leads to two
possible independent solutions for $h(r)$ depending on whether its second-order derivative $h^{(2)}(r)$
vanishes or not. These solutions are respectively
\begin{aleq}\label{eq:4.24}
(\mathrm{i})\ &\quad h(r)=c_1r+c_2,\qquad\quad &&\text{if }h^{(2)}(r)=0,\\
(\mathrm{ii})&\quad h(r)=\frac{c_1}{r+c_2}+c_3,\qquad &&\text{if }h^{(2)}(r)\neq 0,
\end{aleq}%
and their complex conjugates, where the $c_i$ are integration constants. Consider two separate cases: when the function $h$ is given by (\ref{eq:4.24}.i) and when $h$ is of the form
(\ref{eq:4.24}.ii).
%######################### case i #####################
\paragraph{Case i.} In this case, the real
solution of the original system (\ref{eq:4.1}) takes the form
\begin{aleq}\label{eq:4.25}
u(t,x,y)=&4\pa{\mathcal{R}e(c_1(t))x+\mathcal{I}m(c_1(t))y+\mathcal{R}e(c_2(t))},\\
v(t,x,y)=&4\pa{\mathcal{I}m(c_1(t))x-\mathcal{R}e(c_1(t))y+\mathcal{I}m(c_2(t))},\\
\theta(t,x,y)=& \left\{\begin{aligned}
&-\frac{1}{2}\arctan\pa{\frac{\mathcal{R}e(c_1(t))}{\mathcal{I}m(c_1(t))}},\quad
&&\text{if }\mathcal{I}m(c_1(t))\neq0,\\
&\frac{\pi}{4} && \text{if } \mathcal{I}m(c_1(t))=0,
\end{aligned}\right.\\
\sigma(t,x,y)=&-\rho V(t,x,y)+\rho\pa{2|c_1(t)|^2+\mathcal{R}e(\dot{c}_1(t))}x^2
-2\rho\mathcal{I}m(\dot{c}_1(t))xy\\
&+\rho\pa{2|c_1(t)|^2-\mathcal{R}e(\dot{c}_1(t))}y^2+2\rho\pa{2\mathcal{R}e\pa{c_1(t)\bar{c}_2(t)}+\mathcal{R}e(\dot{c}_2(t))}x\\
&-2\rho\pa{2\mathcal{I}m\pa{c_1(t)\bar{c}_2(t)}+\mathcal{I}m(\dot{c}_2(t))}y+\sigma_0(t),
\end{aleq}%
where the real function $\sigma_0(t)$ and the complex functions $c_1(t)$, $c_2(t)$ are arbitrary
functions of time and $\dot{c}_1(t)$, $\dot{c}_2(t)$ are their respective derivatives with respect to
$t$. It should be noted that if the potential $V$, the real function $\sigma_0(t)$ and the complex
functions $c_1(t)$, $c_2(t)$, are all bounded and their derivatives $\dot{c}_1(t)$,
$\dot{c}_2(t)$ are also bounded, then the solution (\ref{eq:4.25}) is bounded. Moreover, if we take
these functions to be of the form
\begin{eqe}\label{eq:4.25b}
\sigma_0(t)=a_0e^{-s_0t},\quad c_i(t)=a_je^{-s_jt}+ib_je^{-q_jt},\quad j=1,2,
\end{eqe}%
where $0<q_j,s_j\in \RR$, $a_j,b_j\in \RR$, $j=0,1,2$, then we obtain a damped solution. Another
physically interesting situation occurs when the arbitrary functions $c_i(t)$ are constant in
solution (\ref{eq:4.25}). In this case, we obtain a stationary solution for the velocities $u$ and $v$.
This allows us to draw the shape of those extrusion dies which are admissible by requiring that the
tool walls coincide with the lines of flow generated by the velocities $u$ and $v$. This is
necessary since we suppose that the flow is incompressible. From the practical point of view, it is
convenient to press the plastic material through the die rectilinearly with constant speed. If we
feed the die in this way, the plasticity region limits are curves defined by the ODE \cite{Lamothe:2}
\begin{eqe}\label{eq:4.26}
\frac{dy}{dx}=\frac{V_0-v(x,y)}{U_0-u(x,y)},
\end{eqe}%
where $U_0$ and $V_0$ are constants representing the feeding velocity (or extraction velocity) of
the die along the $x$-axis and $y$-axis respectively. Equation (\ref{eq:4.26}) is a consequence of
the hypothesis that the flow is incompressible and the mass is conserved. This condition reduces to
the boundary conditions described in \cite{Czyz:1974} when we require that the limits of the
plasticity region correspond to the slip lines (which correspond to the characteristic curves of
the original system (\ref{eq:4.1})), that is, when we require that $dy/dx=\tan\theta$ or
$dy/dx=-\cot\theta$. Here, we use the weakened condition (\ref{eq:4.26}) because no constraints are
imposed on the flow lines which are in contact with the tool walls. This is so because, for a given
solution and given parameters, we choose the walls of the extrusion die to lie along the flow
lines. For the purpose of illustrating the applicability of the method, we have drawn in figure
\ref{fig:1} the shape of a tool and the flow of matter inside the extrusion die for the following
parameters $\mathcal{R}e(c_1(t))=1$, $\mathcal{I}m(c_1(t))=0$, $\mathcal{R}e(c_2(t))=0$,
$\mathcal{I}m(c_2(t))=0$. The feeding velocity of the tool is $U_0=5.95$, $V_0=0$ and the tool
expels the matter at a velocity of $U_1=24.05$, $V_1=0$. The plasticity region at the opening is
bounded by the curve $C_1$ and at the exit by the curve $C_2$. This extrusion die can thin a plate or
rod of ideal plastic material.

\begin{figure}
\begin{center}
\includegraphics[width=3.5in]{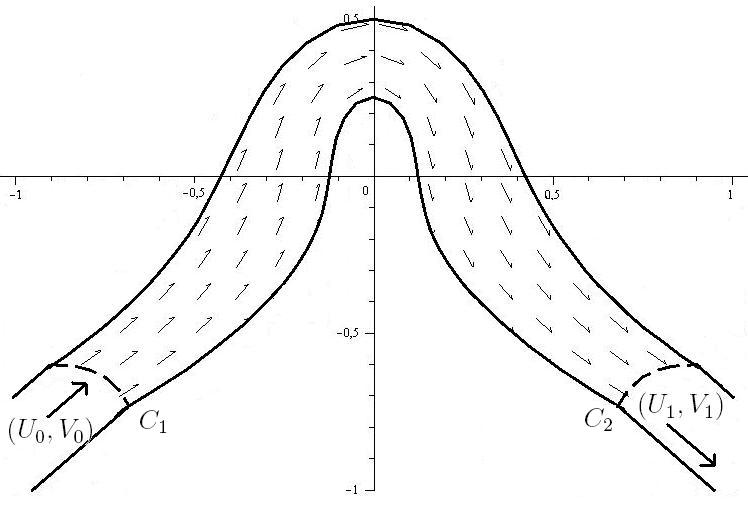}
\end{center}
\caption{Extrusion die corresponding to the solution (\ref{eq:4.27}).} \label{fig:2}
\end{figure}

%###################### Case ii ################################
\paragraph{Case ii.}If $h$ is defined
by (\ref{eq:4.24}.ii), then the corresponding nontrival solution of system (\ref{eq:4.1}) takes the form
\begin{aleq}\label{eq:4.27}
u(t,x,y)=&\mathcal{R}e(c_3(t))+2\frac{\mathcal{R}e\pa{c_1(t)\bar{c}_2(t)}+\mathcal{R}e(c_1(t))x
+\mathcal{I}m(c_1(t))y}{x^2+y^2+2\pa{\mathcal{R}e(c_2(t))x+\mathcal{I}m(c_2(t))y}+|c_2(t)|^2},\\
v(t,x,y)=&\mathcal{R}e(c_3(t))+2\frac{-\mathcal{I}m\pa{c_1(t)\bar{c}_2(t)}-\mathcal{I}m(c_1(t))x
+\mathcal{R}e(c_1(t))y}{x^2+y^2+2\pa{\mathcal{R}e(c_2(t))x+\mathcal{I}m(c_2(t))y}+|c_2(t)|^2},\\
\theta(t,x,y)=&-\frac{1}{2}\arctan\bigg(\bigg[-\mathcal{I}m(c_1(t))(x^2-y^2)+2\mathcal{R}e(c_1(t))xy\\
&+\big(-\mathcal{I}m(c_1(t)\bar{c}_2(t))
+\mathcal{R}e(c_1(t))\mathcal{R}e(c_2(t))-\mathcal{I}m(c_1(t))\mathcal{R}e(c_2(t))\big)x\\
&+\big(\mathcal{R}e(c_1(t)\bar{c}_2(t))
+\mathcal{R}e(c_1(t))\mathcal{R}e(c_2(t))+\mathcal{I}m(c_1(t))\mathcal{R}e(c_2(t))\big)y\\
&+\mathcal{R}e\pa{c_1(t)\bar{c}_2(t)}\mathcal{R}e(c_2(t))
+\mathcal{I}m\pa{c_1(t)\bar{c}_2(t)}\mathcal{I}m(c_2(t))\bigg]^{-1}\\
&\times\bigg[\mathcal{R}e(c_1(t))(x^2-y^2)+2\mathcal{I}m(c_1(t))xy\\
&+\big(-\mathcal{R}e(c_1(t)\bar{c}_2(t))
+\mathcal{R}e(c_1(t))\mathcal{R}e(c_2(t))+\mathcal{I}m(c_1(t))\mathcal{I}m(c_2(t))\big)x\\
&+\big(\mathcal{I}m(c_1(t)\bar{c}_2(t))
-\mathcal{R}e(c_1(t))\mathcal{I}m(c_2(t))+\mathcal{I}m(c_1(t))\mathcal{R}e(c_2(t))\big)y\\
&+\mathcal{R}e\pa{c_1(t)\bar{c}_2(t)}\mathcal{R}e(c_2(t))
+\mathcal{I}m\pa{c_1(t)\bar{c}_2(t)}\mathcal{I}m(c_2(t))\bigg]\bigg),
\end{aleq}%
where the mean pressure $\sigma$ is given by (\ref{eq:4.3}), where we substitute the values of
functions $u,v,\theta$ given by (\ref{eq:4.27}). The complex functions $c_i(t)$, $i=1,2,3$, and the
real function $\sigma_0(t)$ which appear in this solution are arbitrary. For any time $t_0$,
the solution (\ref{eq:4.3}), (\ref{eq:4.27}) has a singularity at point $(x_0,y_0)$ which satisfies the
equation
$$x_0^2+y_0^2+2\pa{\mathcal{R}e(c_2(t_0))x_0+\mathcal{I}m(c_2(t_0))y_0}+|c_2(t_0)|^2=0.$$%
This singularity is stationary if the function $c_2(t)$ is constant. Otherwise, its
position varies with time. If the functions $\sigma_0$ and $c_i$, $i=1,2,3$, are of the form
(\ref{eq:4.25}) defined in a region of the $xy$-plane on a time interval $[T_0,T)$ where the
gradient catastrophe does not occur, then the solution is bounded and damped. In figure
\ref{fig:2}, we have drawn the shape of an extrusion die corresponding to the solution
(\ref{eq:4.27}) for the following choice of parameters: $\mathcal{R}e\pa{c_1(t)}=0$,
$\mathcal{I}m\pa{c_1(t)}=0$, $\mathcal{R}e\pa{c_2(t)}=0$, $\mathcal{I}m\pa{c_2(t)}=-0.5$,
$\mathcal{R}e\pa{c_3(t)}=-0.5$ and $\mathcal{I}m\pa{c_3(t)}=0$. The feeding velocity has component
$U_0=0.2$, $V_0=0.2$, and the extraction of material is performed at the velocity $U_1=0.2$,
$V_1=-0.2$. This type of tool can be used to bend a rod by extrusion without having to fold it. Finally, we should emphasize that the flow changes considerably when the parameters are
varied and we have the freedom to choose the walls of the tool among the flow lines for certain
fixed parameters $c_i(t)$. Moreover, the velocity and the orientation of the feeding (extraction)
can vary somewhat for a given shape of the tool. Consequently, many types of extrusion dies can be
drawn.
%################################## final remarks #######################################
\section{Final remarks}\label{sec:8}The generalized method of characteristics was originally devised for
solving first-order quasilinear hyperbolic systems. The proposed techniques described in Sections \ref{sec:3}, \ref{sec:5} and \ref{sec:6} allow us to extend the
applicability of this approach not only to hyperbolic systems but also to encompass elliptic and
mixed (parabolic) type systems, both homogeneous and inhomogeneous. A variant of the conditional symmetry method for obtaining
multimode solutions has been proposed for these types of systems. We have demonstrated the
usefulness of this approach through the examples of the nonlinear interaction of waves and particles and of the ideal plasticity in $(2+1)$ dimensions in its elliptic region.
New classes of real solutions have been constructed in closed form, some of them bounded. Some of
the obtained solutions describe a stationary flow for an appropriate choice of parameters. For
these solutions, we have drawn extrusion dies and the vector fields which define the flow inside a
region where the gradient catastrophe does not occur.
\paragraph{}The proposed approach for constructing multimode solutions can be used in several potential applications arising from systems describing nonlinear phenomena in physics. It should be noted that in
the multidimensional case, for many physical models, there are few known examples of multimode
solutions written in terms of Riemann invariants for elliptic systems. This is a motivating factor for the elaboration of the generalization of the methods presented in Sections \ref{sec:5} and \ref{sec:6} through the introduction of rotation matrices in the factorization of the Jacobian matrices (\ref{eq:ms:2}). This fact weakens the integrability condition required in the expression (\ref{eq:ms:12}). The approach proposed in this paper offers a new and promising way to construct and investigate such types of solutions. This makes our approach attractive since it can widen the
potential range of applications leading to more diverse types of solutions.
\section*{Acknowledgements}
This work was supported by a research grant from the Natural Sciences and Engineering Council of
Canada (NSERC).

\section*{Appendix: Derivation of the Jacobi Matrix (\ref{eq:3.7})}
\setcounter{equation}{0}
\renewcommand{\theequation}{A.\arabic{equation}}
In the appendix we present some details of the derivation of the equations (\ref{eq:3.7}). We use the notation introduced in the main text. Denote the $q\times q$ matrix
\begin{eqe}\label{eq:A:1}
\psi=I_q-\pa{\pa{\frac{\del f}{\del r}+\frac{\del \bar{f}}{\del r}}\frac{\del r}{\del u}+\text{c.c.}}
\end{eqe}%
Consequently, the expression for the Jacobian matrix (\ref{eq:3.6}) becomes
\begin{eqe}\label{eq:A:2}
\psi^{-1}\pa{\pa{\frac{\del f}{\del r}+\frac{\del \bar{f}}{\del r}}\lambda+\text{c.c.}}.
\end{eqe}%
The expression (\ref{eq:A:1}) can also be written as
\begin{eqe}\label{eq:A:3}
\psi=I_q-F S,
\end{eqe}%
where the matrices $F$ and $S$ are defined in terms of the block matrices ${\del f}/{\del r}+{\del \bar{f}}/{\del r}\in\RR^{q\times k}$ and ${\del r}/{\del u}\in\RR^{k\times q}$ together with their complex conjugates
\begin{gaeq}\label{eq:A:4}
F=\pa{
       \frac{\del f}{\del r}+\frac{\del \bar{f}}{\del r}, \frac{\del f}{\del \bar{r}}+\frac{\del \bar{f}}{\del \bar{r}}
}\in\RR^{q\times 2k},\\
S=\pa{\begin{array}{c}
        \frac{\del r}{\del u}\\
        \frac{\del \bar{r}}{\del u}\\
      \end{array}
}\in\RR^{2k\times q}.
\end{gaeq}%
We multiply equation (\ref{eq:A:3}) on the right by the matrix $F$ in order to obtain
\begin{eqe}
\psi F=F-FSF=F(I_{2k}-S F)=F \Phi,
\end{eqe}%
where
\begin{eqe}\label{eq:A:4:b}
\Phi=I_{2k}-SF.
\end{eqe}%
Assuming that $\psi$ and $\Phi$ are invertible, the relation
\begin{eqe}\label{eq:A:5}
\psi^{-1}F=F\Phi^{-1}
\end{eqe}%
is satisfied. Using the notations (\ref{eq:A:3}) and (\ref{eq:A:4}), the Jacobian matrix (\ref{eq:A:2}) can be written in the compact form
\begin{eqe}\label{eq:A:6}
\del u=\psi^{-1}F\Lambda,\quad \Lambda=\pa{\lambda,\bar{\lambda}}^{T}.
\end{eqe}%
Replacing (\ref{eq:A:5}) into (\ref{eq:A:6}), we obtain the expression for the Jacobian matrix
\begin{eqe}\label{eq:A:7}
\del u=F\Phi^{-1}\Lambda.
\end{eqe}%
We express the inverse of the matrix
\begin{eqe}\label{eq:A:8}
\Phi=\pa{\begin{array}{cc}
           \mathcal{A} & \mathcal{B} \\
           \bar{\mathcal{B}} & \bar{\mathcal{A}}
         \end{array}
}
\end{eqe}%
in terms of the block matrices
\begin{eqe}\label{eq:A:9}
\mathcal{A}=I_k-\frac{\del r}{\del u}\pa{\frac{\del f}{\del r}+\frac{\del \bar{f}}{\del r}},\quad \mathcal{B}=\frac{\del r}{\del u}\pa{\frac{\del f}{\del \bar{r}}+\frac{\del \bar{f}}{\del \bar{r}}},
\end{eqe}%
where $\bar{\mathcal{A}}$ and $\bar{\mathcal{B}}$ are the complex conjugates of $\mathcal{A}$ and $\mathcal{B}$ respectively. Assuming that the block $\mathcal{A}$ is invertible, the matrix $\Phi$ is also invertible when
$$\det\Phi=\det\pa{\mathcal{A}}\pa{\bar{\mathcal{A}}-\bar{\mathcal{B}}\mathcal{A}^{-1}\mathcal{B}}\neq 0.$$%
Consequently, we require that
\begin{eqe}\label{eq:A:10}
\det\mathcal{A}\neq0,\qquad \bar{\mathcal{A}}-\bar{\mathcal{B}}\mathcal{A}^{-1}\mathcal{B}\neq 0.
\end{eqe}%
It should be noted that for $r$ and $f(r,\bar{r})$ of class $C^2$, the matrix $\mathcal{A}$ is the identity $I_k$ in the neighborhood of $x=0$, and it has to be invertible. Under the assumption (\ref{eq:A:10}), the inverse of the matrix $\Phi$ is given by
\begin{eqe}\label{eq:A:11}
\Phi^{-1}=\pa{\begin{array}{cc}
                M^1 & M^2 \\
                \bar{M}^1 & \bar{M}^2
              \end{array}
},
\end{eqe}%
where
\begin{eqe}\label{eq:A:12}
M^1=\pa{\mathcal{A}-\mathcal{B}\bar{\mathcal{A}}^{-1}\bar{\mathcal{B}}}^{-1},\quad M^2=-M^1\mathcal{B}\bar{\mathcal{A}}^{-1}.
\end{eqe}%
Indeed,
\begin{aleq}
\pa{\begin{array}{cc}
                M^1 & M^2 \\
                \bar{M}^1 & \bar{M}^2
              \end{array}
}=&\pa{\begin{array}{cc}
                M^1\mathcal{A}+M^2\mathcal{B} & \quad M^1\mathcal{B}+M^2\bar{\mathcal{A}} \\
                \bar{M}^2\mathcal{A}+\bar{M^2}\mathcal{B} & \quad \bar{M}^2\mathcal{B}+\bar{M}^1\bar{\mathcal{A}}
              \end{array}
}\\
=&\pa{\begin{array}{cc}
                M^1(\mathcal{A}-\mathcal{B}\bar{\mathcal{A}}^{-1}\bar{\mathcal{B}}) & M^1(\mathcal{B}-\mathcal{B}\bar{\mathcal{A}}^{-1}\bar{\mathcal{A}}) \\
                \bar{M}^1(-\bar{\mathcal{B}}\mathcal{A}^{-1}\mathcal{A}+\bar{\mathcal{B}}) & \bar{M}^1(-\mathcal{B}\mathcal{A}^{-1}\mathcal{B}+\bar{\mathcal{A}})
              \end{array}
}\\
=&\pa{\begin{array}{cc}
        I_k & 0 \\
        0 & I_k
      \end{array}
}=I_{2k}.
\end{aleq}%
Replacing the expression (\ref{eq:A:11}) for the matrix $\Phi^{-1}$ into equation (\ref{eq:A:7}), we obtain the form (\ref{eq:3.7}) for the Jacobian matrix. Next, substituting the equations (\ref{eq:A:9}) into the conditions (\ref{eq:A:10}) and into the matrices (\ref{eq:A:12}), we find the inequalities (\ref{eq:3.11}) and the equations (\ref{eq:3.7b}), respectively.

\bibliographystyle{alpha}% plain,unsrt,alpha

\begin{thebibliography}{10}
\bibitem{GrundlandVassiliou:1991}
A.M. Grundland, P. Vassiliou,
\newblock On the solvability of the Cauchy problem for the Riemann double waves by the Monge-Darboux method.
\newblock {\em Analysis 11}, 221-278, (1991).
\end{thebibliography}

\end{document}